\documentclass[10pt]{article}
\usepackage{amsmath}
\usepackage{amssymb}
\usepackage{graphicx}
\usepackage{color}
\usepackage{orcidlink}
\usepackage{hyperref}
\usepackage{setspace}
\hypersetup{colorlinks=true, linkcolor=blue, citecolor=blue, urlcolor=blue}
\usepackage{caption}
\usepackage{subcaption}
\begin{document}
\baselineskip=16pt
\onehalfspacing

\begin{center}
\LARGE{Thermodynamics of BTZ-type charged black holes in Bopp Podolsky electrodynamics}
\end{center}

\vspace{0.3cm}

\begin{center}

{\bf Allan. R. P. Moreira\orcidlink{0000-0002-6535-493X}}\footnote{\bf allan.moreira@fisica.ufc.br }\\
\vspace{0.1cm}
{\it Research Center for Quantum Physics, Huzhou University, Huzhou 313000, China}\\
{\it Secretaria da Educa\c{c}\~{a}o do Cear\'{a} (SEDUC), Coordenadoria Regional de Desenvolvimento da Educa\c{c}\~{a}o (CREDE 9),  Horizonte, Cear\'{a}, 62880-384, Brazil}\\
\vspace{0.2cm}
{\bf Abdelmalek Bouzenada\orcidlink{0000-0002-3363-980X}}\footnote{\textbf{abdelmalekbouzenada@gmail.com (Corresp. author) }}\\
\vspace{0.1cm}
{\it Laboratory of Theoretical and Applied Physics, Echahid Cheikh Larbi Tebessi University 12001, Algeria}\\
\vspace{0.2cm}
{\bf Guo-Hua Sun\orcidlink{0000-0002-0689-2754}}\footnote{\bf sunghdb@yahoo.com }\\
\vspace{0.1cm}
{\it Centro de Investigaci\'{o}n en Computaci\'{o}n, Instituto Polit\'{e}cnico Nacional, UPALM, CDMX 07700, Mexico}\\
\vspace{0.2cm}
{\bf Shi-Hai Dong\orcidlink{0000-0002-0769-635X}}\footnote{\bf dongsh2@yahoo.com }\\
\vspace{0.1cm}
{\it Research Center for Quantum Physics, Huzhou University, Huzhou 313000, China}\\
{\it Centro de Investigaci\'{o}n en Computaci\'{o}n, Instituto Polit\'{e}cnico Nacional, UPALM, CDMX 07700, Mexico}\\
\vspace{0.2cm}

\end{center}

\vspace{0.2cm}

\begin{abstract}

We tested the thermodynamic properties of charged BTZ-type black holes (BHs) in the framework of Bopp-Podolsky electrodynamics, a higher-derivative extension of Maxwell’s theory that preserves gauge invariance while introducing a massive photon mode. Using a perturbative approach, we derive first- and second-order corrections to the metric and electric field, revealing how the Bopp-Podolsky parameter $ b $ modifies the geometry and horizon structure. Unlike in four-dimensional illustrations, where such corrections can lead to wormhole solutions, the (2+1)-dimensional case retains a black hole interpretation, albeit with curvature-dependent deformations that vanish asymptotically. We compute the Hawking temperature via the Hamilton-Jacobi tunneling method, demonstrating its dependence on both the cosmological constant and the Bopp-Podolsky coupling. Our results indicate that while energy conditions are marginally violated for certain parameter regimes, the BHs thermodynamic behavior remains well-defined, with temperature corrections emerging from the interplay between higher-derivative electrodynamics and the lower-dimensional gravitational background.
\end{abstract}

\vspace{0.1cm}


\vspace{0.1cm}


%%%%%%%%%%%%%%%%%%%%%%%%%%%%%%%%%%%%%%%%%%%%%%%%%%%%%%%%%%%%%%%%%%%%%%%%%%%%%%%%%%%%%%
\section{Introduction}\label{intro}

In these theories, Bopp \cite{1940} and Podolsky \cite{Podolsky1942} sought to expalin the infinities arising in the conventional treatment of point charges by introducing higher-order derivatives ($\mathcal{D}$) into the Lagrangian of electrodynamics while preserving linearity in the equations of motion and maintaining gauge invariance. Podolsky, in particular, tested the classical aspects of his model, deriving the equations of motion, energy-momentum tensor, and plane wave solutions~\cite{Podolsky1942}. Over time, the Bopp-Podolsky (BP) model has been widely interpreted as a framework for describing massive photons without violating gauge invariance, as its propagating modes include both massless photons and massive ones. This work demonstrates that the BP model's solution for a point charge in electrodynamics can be reinterpreted as an ordinary electrodynamic solution corresponding to a specific charge distribution. Building on this insight, we further clarify how the BP model naturally provides Pauli-Villars (PV) regularization in standard QED, a connection previously noted by Kvasnica~\cite{Kvasnica1960}, who also identified the charge distribution but treated the BP model's mass parameter as a physical mass to be compared with Hoffstadter’s scattering experiments~\cite{Hoffstadter1956}. Additionally, in the Lorenz gauge, a gauge-fixing term introduced by Podolsky and Kikuchi~\cite{PodolskyKikuchi1944}, which simplifies quantization calculations, had been overlooked until recently revisited by Bufalo, Pimentel, and Soto~\cite{Bufalo2014}. In Bopp-Podolsky electrodynamics, the standard Lorenz gauge condition is replaced by a modified version~\cite{GalvaoPimentel1988} that better accommodates the five degrees of freedom in the spectrum, two for the massless photon mode and three for the massive longitudinal mode~\cite{Ferreira2019}. Research on BP electrodynamics has expanded into various areas, including low-energy radiative corrections~\cite{Borges2019}, renormalization~\cite{Bufalo2012}, path integral quantization~\cite{Bufalo2011}, finite-temperature effects~\cite{Bonin2010, AraujoFilho2021}, multipole expansions~\cite{Bonin2019}, black hole and wormhole solutions~\cite{Cuzinatto2018, Frizo2023}, cosmological applications~\cite{Cuzinatto2017}, and numerous other contexts~\cite{Kruglov2010, Cuzinatto2011, Zayats2014, Granado2020}, shown its broad theoretical and phenomenological relevance.

More recently, interest has grown in examining the implications of Bopp–Podolsky electrodynamics in curved backgrounds, especially concerning the no-hair conjecture. According to this conjecture, or theorem in more restrictive formulations \cite{Israel:1967wq,Israel:1967za,Carter:1971zc}, stationary black hole solutions in general relativity coupled to standard Maxwell fields are entirely characterized by global charges such as mass, electric charge, and angular momentum. Within this framework, Cuzinatto et al. \cite{Cuzinatto2017} showed that, for spherically symmetric spacetimes, only the massless sector of the Bopp–Podolsky field propagates outside the horizon, effectively respecting the no-hair condition. Nonetheless, the presence of the Bopp–Podolsky parameter $b$ introduces a scale that could, in principle, allow for non-trivial field configurations outside the black hole.

The notion of black holes (BHs) \cite{B1,B2,B3,B4,B5,B6,B7,B8,B9,B10} is often accompanied by the concept of Hawking radiation, which plays a fundamental role in establishing a thermodynamic perspective for these objects \cite{Gomes2018oyd}. This perspective gained structure through the pioneering work of Bardeen and collaborators \cite{Bardeen}, who systematically formulated four laws governing black hole mechanics. These laws were inspired by earlier ideas from Bekenstein \cite{Bekenstein3}, particularly his insights into black hole entropy. The striking resemblance between these mechanical laws and the classical laws of thermodynamics led to the interpretation of black holes as genuine thermodynamic systems, characterized by quantities such as temperature and entropy. Today, this framework is widely accepted, and the laws derived in this context are collectively known as the laws of black hole thermodynamics.

Our work is motivated by understanding the behavior of Bopp-Podolsky electrodynamics in a gravitational background of dimension $(2+1)$. We seek to understand how Bopp-Podolsky corrections modify the perturbative solution analogous to a charged BTZ black hole. Furthermore, we seek to understand the influence of these corrections on the Hawking temperature measurements of the black hole. In what follows, we adopt geometrized units ($G = c = 1$) and work with a metric signature $(+,-,-)$. The implications of the Bopp–Podolsky parameter on the curvature and electromagnetic fields are discussed in detail within this lower-dimensional framework.

\section{Bopp-Podolsky Theory in Curved Spacetime \label{S2}}

We begin by formulating the action and field equations for the Bopp-Podolsky theory in a curved background. In line with the generalized approach proposed in Refs.~\cite{Cuzinatto2017, Frizo2023}, the Lagrangian density incorporates higher-order derivatives of the field strength and non-minimal curvature couplings, thereby extending the conventional Maxwell framework. The modified electrodynamics is governed by the following Lagrangian:
\begin{equation}
\mathcal{L}_m = -\frac{1}{4} F^{\mu\nu}F_{\mu\nu} 
+ \frac{1}{2}(a^2 + 2b^2)\nabla_\nu F^{\mu\nu} \nabla^\rho F_{\mu\rho} 
+ b^2 \left(R_{\mu\nu}F^{\mu\alpha}F^\nu_{\ \alpha} + R_{\mu\nu\alpha\beta}F^{\mu\beta}F^{\nu\alpha} \right),
\label{LagCurvo}
\end{equation}
where $F_{\mu\nu} = \nabla_\mu A_\nu - \nabla_\nu A_\mu$ is the antisymmetric field tensor, $\nabla_\mu$ denotes the covariant derivative, $R_{\mu\nu}$ is the Ricci tensor, and $R_{\mu\nu\alpha\beta}$ is the Riemann tensor. The constants $a$ and $b$ parametrize the strength of higher-derivative and non-minimal contributions, respectively.

The Lagrangian in Eq.~\eqref{LagCurvo} preserves both local Lorentz invariance and the $U(1)$ gauge symmetry. Notably, it features up to fourth-order derivatives in the gauge field and remains quadratic in both $A_\mu$ and its derivatives.

Coupling this extended electrodynamics to gravity in (2+1) dimensions, we define the total action as:
\begin{equation}
S = \frac{1}{16\pi} \int d^3x \sqrt{-g} \left( -R - 2\Lambda + 4\mathcal{L}_m \right),
\label{acaoTotal}
\end{equation}
where $g$ is the determinant of the metric $g_{\mu\nu}$, $R$ is the Ricci scalar, and $\Lambda$ is the cosmological constant.

Varying the action \eqref{acaoTotal} with respect to the metric leads to the modified Einstein equations:
\begin{equation}
R_{\mu\nu} - \frac{1}{2} g_{\mu\nu} R - \Lambda g_{\mu\nu} = 8\pi \left( T^{M}_{\mu\nu} + T^{a}_{\mu\nu} + T^{b}_{\mu\nu} \right),
\label{EinsteinEq}
\end{equation}
where the energy-momentum tensor decomposes into contributions from the Maxwell term ($T^M_{\mu\nu}$), the higher-derivative correction ($T^a_{\mu\nu}$), and the curvature-coupling correction ($T^b_{\mu\nu}$), given respectively by:
\begin{align}
T^{M}_{\mu\nu} &= \frac{1}{4\pi} \left( F_{\mu\rho}F_{\nu}^{\ \rho} + \frac{1}{4}g_{\mu\nu} F^{\alpha\beta}F_{\alpha\beta} \right), \nonumber\\
T^{a}_{\mu\nu} &= \frac{a^2}{4\pi} \left( 
g_{\mu\nu} F_\beta^{\ \gamma} \nabla_\gamma K^\beta + \frac{1}{2}g_{\mu\nu} K^\beta K_\beta 
+ 2 F_{(\mu}^{\ \ \alpha} \nabla_{\nu)} K_\alpha 
- 2 F_{(\mu}^{\ \ \alpha} \nabla_\alpha K_{\nu)} 
- K_\mu K_\nu \right), \nonumber \\
T^{b}_{\mu\nu} &= \frac{b^2}{2\pi} \left( 
-\frac{1}{4} g_{\mu\nu} \nabla^\beta F^{\alpha\gamma} \nabla_\beta F_{\alpha\gamma}
+ F_{(\mu}^{\ \ \gamma} \nabla^\beta \nabla_\beta F_{\nu)\gamma}
+ F_{\gamma(\mu} \nabla_\beta \nabla_{\nu)} F^{\beta\gamma} 
- \nabla_\beta \left(F^\beta_{\ \ \gamma} \nabla_{(\mu} F_{\nu)}^{\ \ \gamma} \right) \right).
\label{Tb}
\end{align}
Here, $K^\mu = \nabla_\nu F^{\mu\nu}$, and parentheses denote symmetrization over indices.

To derive the generalized Maxwell equations, we vary the action with respect to $A_\mu$, obtaining:
\begin{equation}
\nabla_\nu \left[ F^{\mu\nu} - (a^2 + 2b^2) H^{\mu\nu} + 2b^2 S^{\mu\nu} \right] = 0,
\label{EqEM}
\end{equation}
where
\begin{align}
H^{\mu\nu} &= \nabla^\mu K^\nu - \nabla^\nu K^\mu,  \nonumber\\
S^{\mu\nu} &= F^{\mu\lambda} R_\lambda^{\ \nu} - F^{\nu\lambda} R_\lambda^{\ \mu} + 2 R^{\mu\ \nu}_{\ \rho\ \sigma} F^{\sigma\rho}. 
\label{Smn}
\end{align}

The conservation of the total energy-momentum tensor,
\begin{equation}
\nabla_\nu T^{\mu\nu} = 0,
\label{Conservacao}
\end{equation}
is guaranteed by diffeomorphism invariance of the action.

In four-dimensional setups, Cuzinatto et al.~\cite{Cuzinatto2017} obtained analytical insights by applying the Bekenstein method, identifying the Reissner–Nordström solution as a limiting case for $b=0$, and discussing the emergence of hair-like configurations for $b\neq0$. Frizo et al.~\cite{Frizo2023} later applied a perturbative scheme and found that only the $b$-term influences the solution at first order, leading to geometries interpreted as wormholes rather than black holes. They argue that $b$ should be seen as a universal parameter, not a characteristic of the black hole, in line with the no-hair conjecture.

In this work, we extend the perturbative analysis to second order in $(2+1)$ dimensions, seeking corrections around the charged BTZ background. We show that, analogous to the 4D case, only the $b$-parameter contributes at leading orders. However, in contrast to Ref.~\cite{Frizo2023}, the corrected geometry still describes a black hole--specifically, a charged BTZ solution modified by curvature-dependent electrodynamic effects that vanish asymptotically but become relevant near the origin.

To proceed, we adopt the following static and circularly symmetric metric \textit{ansatz}:
\begin{equation}
ds^2 = A(r) dt^2 - \frac{dr^2}{B(r)} - r^2 d\theta^2,
\label{ansatz}
\end{equation}
where $A(r)$ and $B(r)$ are functions to be determined. We assume a purely electric configuration with vanishing magnetic field, such that the field strength tensor simplifies to:
\begin{equation}
F_{\mu\nu} = E(r) (\delta^1_\mu \delta^0_\nu - \delta^0_\mu \delta^1_\nu),
\label{CampoEM}
\end{equation}
with $E(r)$ denoting the radial electric field.

Given the complexity of the field equations, we resort to a perturbative expansion. The functions $A(r)$, $B(r)$, and $E(r)$ are expanded as:
\begin{align}
A(r) &= A_0(r) + \xi A_1(r) + \xi^2 A_2(r) + \mathcal{O}(\xi^3),\nonumber \\
B(r) &= B_0(r) + \xi B_1(r) + \xi^2 B_2(r) + \mathcal{O}(\xi^3), \nonumber \\
E(r) &= E_0(r) + \xi E_1(r) + \xi^2 E_2(r) + \mathcal{O}(\xi^3), \label{PertE}
\end{align}
where $\xi \equiv \xi(a,b)$ is a small perturbation parameter. The zeroth-order background corresponds to the uncharged BTZ solution and the standard Coulomb electric field:
\begin{align}
A_0(r) &= B_0(r) = -M - \Lambda r^2,\nonumber \\
E_0(r) &= \frac{Q}{r}.
\end{align}

In the following, we analyze how the Bopp-Podolsky corrections deform this geometry, emphasizing the role of the $b$ parameter in the modified gravitational field.

\section{Hawking temperature}

To study the BH temperature, we will adopt the Hamilton-Jacobi formalism through the tunneling approach \cite{Srinivasan1998ty,Angheben2005rm,Kerner2006vu,Mitra2006qa,Akhmedov2006pg}.
The quantum tunneling approach offers a semiclassical interpretation for the emission of radiation by black holes, based on the analysis of the probability of particles crossing the event horizon. In this scenario, Hawking radiation is treated as an emission process originated by the spontaneous creation of particle-antiparticle pairs near the horizon. The particle with negative energy is absorbed by the black hole, contributing to its mass decrease, while its positive energy counterpart can emerge to infinity, crossing a potential barrier via quantum tunneling. The tunneling rate can then be related to the temperature of the black hole. One of the strengths of this technique lies in the fact that the extracted thermodynamic features are intrinsically linked to the geometry of spacetime, which makes its application feasible in a wide class of gravitational metrics and solutions \cite{Jiang2006,Kerner2007rr,Ma2014qma,Maluf2018lyu,Gomes2020kyj}.

In the vicinity of the event horizon, the angular part of the metric is redshifted. Consequently, only the temporal and radial parts will have a significant contribution, so that we can treat the metric (\ref{ansatz}) as being two-dimensional
\begin{align}
ds^2=-A(r)dt^2+B(r)^{-1}dr^2,
\end{align}
the angular part is disregarded, as it is shifted towards the red. Let us consider a perturbation of a massive scalar field $\phi$ around the black hole background, i. e.,
 \begin{align}
     \hbar^2g^{\mu\nu}\nabla_\mu\nabla_\nu\phi-m^2\phi=0,
 \end{align}
as it were, the above equation is the Klein-Gordon equation where $m$ is the mass associated with the $\phi$ field. Using the decomposition of spherical harmonics, one arrives at
 \begin{align}\label{0101}
    -\partial_t^2\phi+\Lambda(r)\partial_r^2\phi+\frac{1}{2}\partial_r\Lambda(r)\partial_r\phi-\frac{m^2}{\hbar^2}A(r)\phi=0,
 \end{align}
where $\Lambda(r)\equiv A(r)B(r)$.
Since $\phi$ is a semiclassical field associated with particles created in the black hole, we can use the so-called Wentzel–Kramers–Brillouin approximation \cite{Sakurai} through the ansatz \cite{Srinivasan1998ty,Angheben2005rm,Kerner2006vu,Mitra2006qa,Akhmedov2006pg}
 \begin{align}
     \phi(t,r)=\exp\Big[\frac{1}{\hbar}\mathcal{T}(t,r)\Big],
 \end{align}
to obtain the solution of Eq. (\ref{0101}).

Note that the Eq.(\ref{0101}) for the lowest order in $\hbar$ is
 \begin{align}\label{0102}
 (\partial_t\mathcal{T})^2-\Lambda(r)(\partial_r\mathcal{T})^2-m^2A(r)\phi=0,
 \end{align}
where particle-like solutions have the form \cite{Srinivasan1998ty,Angheben2005rm,Kerner2006vu,Mitra2006qa,Akhmedov2006pg}
\begin{align}\label{0103}
\mathcal{T}(t,r)=-\omega t+W(r),
\end{align}
here, $\omega$ is a constant that can be thought of as the energy of the emitted radiation. By replacing the solution (\ref{0103}) in Eq. (\ref{0102}), it is found that
\begin{align}\label{0104}
W(r)=\pm\int{\frac{dr}{\sqrt{A(r)B(r)}}\sqrt{\omega^2-m^2A(r)}},
\end{align}
where the positive solution represents the output particle, and the negative solution the input particle. Let us focus on the output solution, as they represent particles emitting radiation as they cross the event horizon. Expanding $A(r)$ and $B(r)$ around the event horizon radius $r_H$
\begin{align}
A(r)=A(r_H)+A'(r_H)(r-r_H)+...\nonumber\\
B(r)=B(r_H)+B'(r_H)(r-r_H)+...
\end{align}
where the prime $'$ denotes differentiation concerning $r$.

In this way, Eq. (\ref{0104}) then becomes
\begin{align}\label{0105}
W(r)=\int{\frac{dr}{\sqrt{A'(r_H)B'(r_H)}}\frac{\sqrt{\omega^2-m^2A'(r_+)(r-r_+)}}{(r-r_+)}}.
\end{align}

Using the residue theorem to solve the integral (\ref{0105}), we obtain that
\begin{align}
W(r_{H})=\frac{2\pi i \omega}{\sqrt{A'(r_H)B'(r_H)}}+(\text{real contribution}).
\end{align}

Now, we return to $\mathcal{T}(t,r)$, whose imaginary part can be related to the tunneling probability of a particle with energy $\omega$ \cite{Srinivasan1998ty,Angheben2005rm,Kerner2006vu,Mitra2006qa,Akhmedov2006pg}. Thus, we are left with 
\begin{align}\label{0106}
\Gamma\sim\exp(-2im\mathcal{T})=\exp\Big[-\frac{4\pi\omega}{\sqrt{A'(r_H)B'(r_H)}}\Big].
\end{align}

Finally, we can identify this probability with the Boltzmann factor  $\Gamma \sim e^{-\beta\omega}$, where $\beta=1/T_{BH}$ (we made $k_B = 1$) to obtain the temperature of the black hole
\begin{align}\label{0107}
T_{BH}=\frac{\omega}{2im\mathcal{T}}=\frac{\sqrt{A'(r_H)B'(r_H)}}{4\pi}.
\end{align}

\subsection{First Approach}

For our BH model, we can initially consider as a solution the first order approximation of Eq.(\ref{PertE})
\begin{align}
 \frac{1}{2r}B_{1}'(r)+&\frac{Q^{2}}{r^{2}(M+\Lambda r^{2})}\left(A_{1}(r)-B_{1}(r)\right)+\frac{2Q}{r}E_{1}(r)\nonumber \\+&\frac{Q^{2}}{\xi r^{4}}\big(r^{2}+4b^{2}(M-\Lambda r^{2})\big) =0,\nonumber \\
 \frac{1}{2r}A_{1}'(r)+&\frac{(Q^{2}-\Lambda r^{2})}{r^{2}(M+\Lambda r^{2})}\left(A_{1}(r)-B_{1}(r)\right)+\frac{2Q}{r}E_{1}(r)\nonumber \\ +&\frac{Q^{2}}{\xi r^{4}}\left(r^{2}-4b^{2}(M+3\Lambda r^{2})\right)=0, \nonumber \\
 \frac{1}{2}A_{1}''(r)-&\frac{\Lambda r}{2(M+\Lambda r^{2})}\left(A_{1}'(r)-B_{1}'(r)\right)-\frac{2Q}{r}E_{1}(r) \nonumber \\ -&\frac{\big(MQ^{2}+(M+Q^{2})\Lambda r^{2}\big)}{r^{2}(M+\Lambda r^{2})^{2}}\left(A_{1}(r)-B_{1}(r)\right)  -\frac{Q^{2}}{\xi r^{4}}\left(r^{2}-4b^{2}(3M+\Lambda r^{2})\right)=0.  
\end{align}
In this approximation regime, all terms containing factors such as $\xi a^2$, $\xi b^2$ or higher powers of $\xi$ are disregarded, since such contributions are subdominant. Analogously to what occurs in the $(3+1)$-dimensional case, terms proportional to $a^2$ appear only at higher orders, typically in $\mathcal{O}(\xi^2)$, and are therefore neglected based on the assumption that $a^2$ and $b^2$ represent small parameters. Furthermore, in this expansion order, no terms with higher-order derivatives appear, and the electric field is obtained directly, without the presence of spatial or temporal derivatives. Therefore, we obtain
 \begin{align}
 ds^2=& \Big[-M-\Lambda r^{2}+\Sigma_1(r)\Big] dt^2 -\frac{dr^2}{\Big[-M-\Lambda r^{2}+\Omega_1(r)\Big]}
  -r^2 d\theta^2,\nonumber \\
  E_1(r)=&\frac{Q}{r}\left(1+2b^2\Lambda\right)+\frac{4b^2 Q^3}{r^3},
\label{Metric1}
 \end{align}
\begin{figure}[ht]
\centering
\includegraphics[height=5cm,width=7cm]{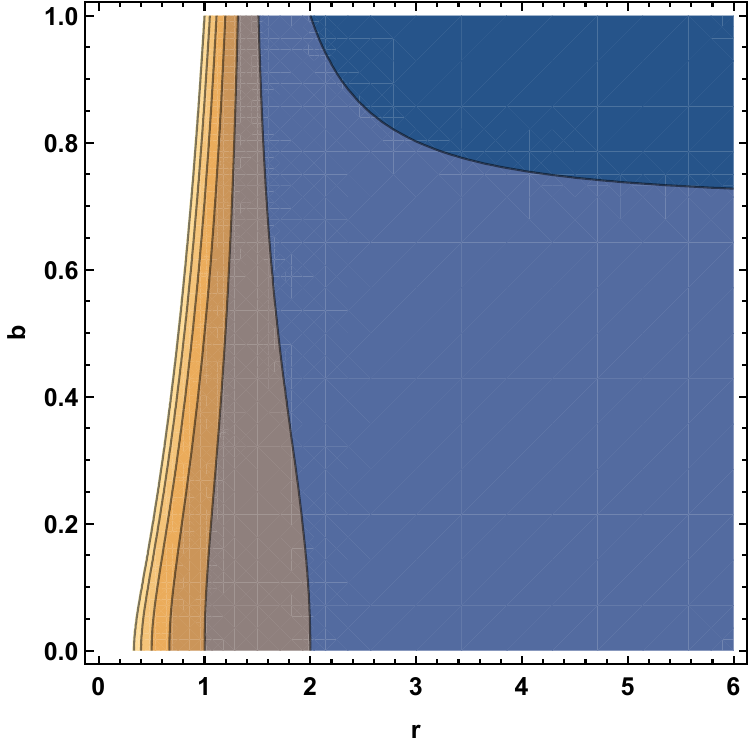}\\
(a)\\
\includegraphics[height=5cm,width=7cm]{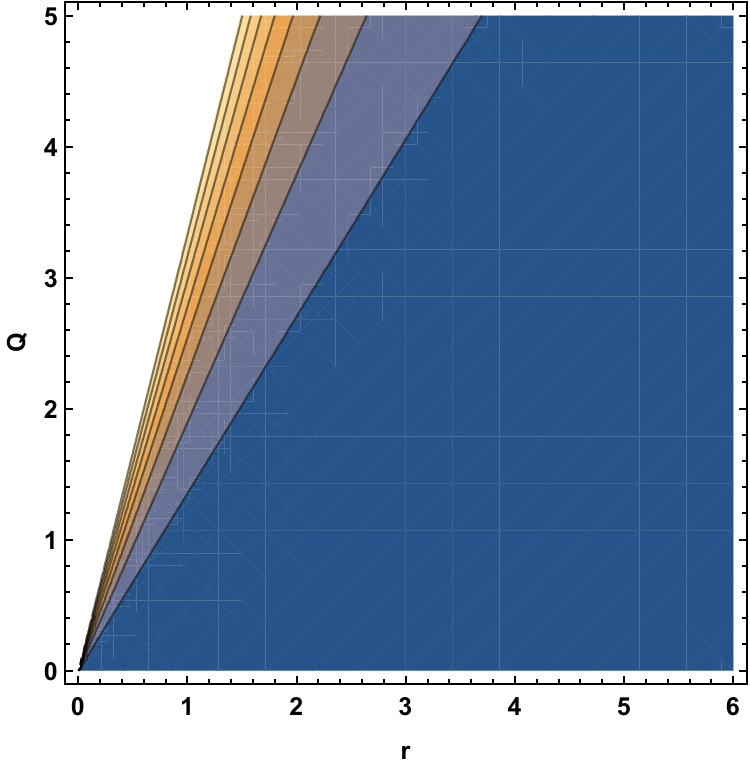}
\includegraphics[height=5cm,width=7cm]{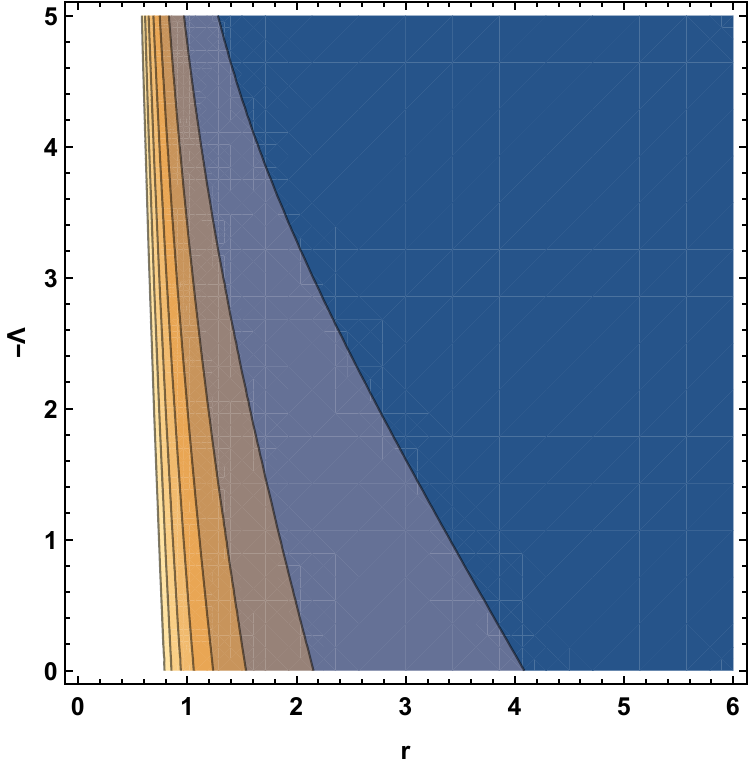}\\
(b) \hspace{8cm} (c)
\caption{Electric field. (a) $\Lambda=-1$ and $Q=M=r_0=1$. (b) $\Lambda=-1$, $b=0.1$ and $M=r_0=1$. (c) $b=0.1$ and $Q=M=r_0=1$.}
\label{fig1}
\end{figure}
where 
 \begin{align}
\Sigma_1(r)&=-\frac{4b^{2}Q^{2}M}{r^{2}} -2Q^{2}\ln\left(\frac{r}{r_{0}}\right),\nonumber \\
 \Omega_1(r)&=8b^2Q^2\Lambda+\frac{4b^{2}Q^{2}M}{r^{2}} -2Q^{2}\ln\left(\frac{r}{r_{0}}\right),
 \end{align}
 and $r_0$ is a positive scale parameter.

It is noted that the metric functions $A(r)$ and $B(r)$ do not coincide, unlike what occurs in spherically symmetric solutions of general relativity with electric charge, where one usually has $A(r) = B(r)$. This type of asymmetry between the temporal and radial components of the metric is typical in gravitational theories that incorporate non-minimal couplings between vector fields and curvature, as is the case of certain models with spontaneous breaking of Lorentz symmetry.

Figure (\ref{fig1}) shows the behavior of the electric field $E(r)$ in the presence of Bopp–Podolsky electrodynamics for a charged BTZ-type BH, analyzed under three distinct parameter sets. In panel (a), the electric field is plotted for different values of the Bopp–Podolsky parameter $b$, with the parameters fixed as $\Lambda = -1$ and $Q = M = r_0 = 1$. The plot shows that as $b$ increases, the electric field becomes more intense near the origin, reflecting the influence of higher-derivative corrections. In panel (b), the variation of the electric field concerning the electric charge $Q$ is examined, with fixed parameters $\Lambda = -1$, $b = 0.1$, and $M = r_0 = 1$, the field grows noticeably stronger with increasing $Q$. Also, panel (c) tests the influence of the cosmological constant $\Lambda$, using fixed values $b = 0.1$ and $Q = M = r_0 = 1$. It reveals that increasing $|\Lambda|$ slightly alters the field profile, especially in the near-horizon region.

\begin{figure}[ht]
\centering
\includegraphics[height=5cm,width=7cm]{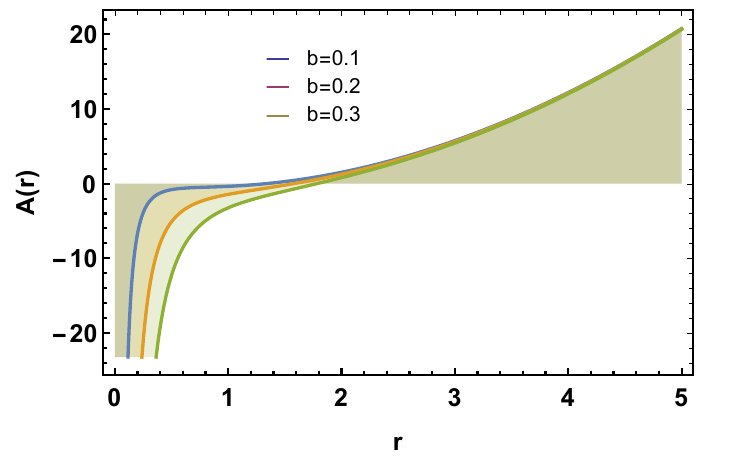}\\
(a)\\
\includegraphics[height=5cm,width=7cm]{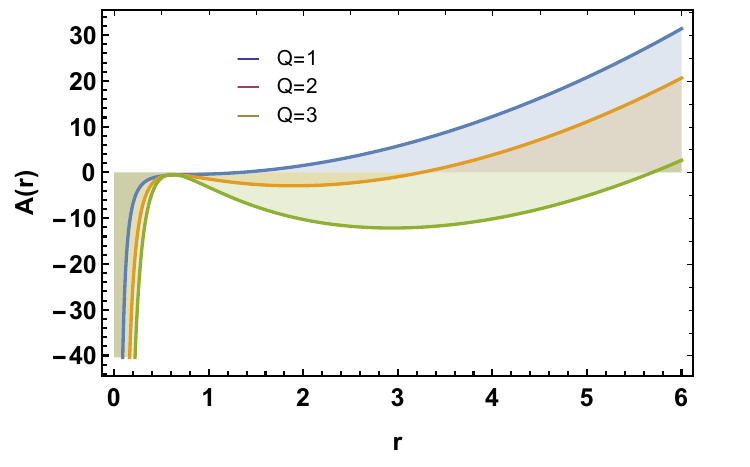}
\includegraphics[height=5cm,width=7cm]{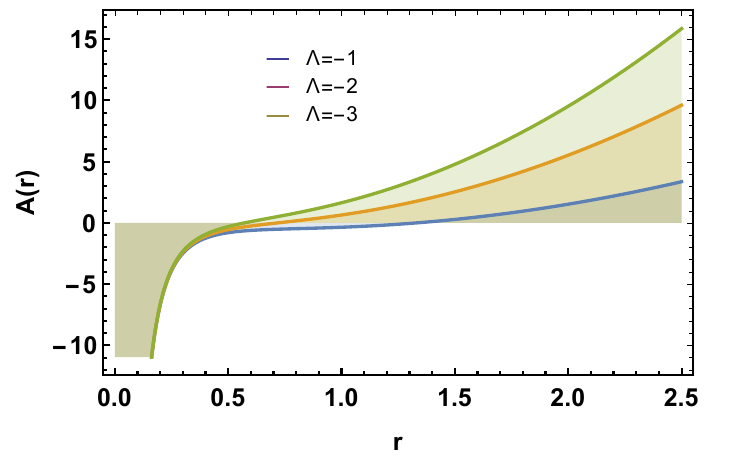}\\
(b) \hspace{8cm} (c)
\caption{Metric coefficient $A(r)$.(a) $\Lambda=-1$ and $Q=M=r_0=1$. (b) $\Lambda=-1$, $b=0.1$ and $M=r_0=1$. (c) $b=0.1$ and $Q=M=r_0=1$.}
\label{fig2}
\end{figure}
\begin{figure}[ht]
\centering
\includegraphics[height=5cm,width=7cm]{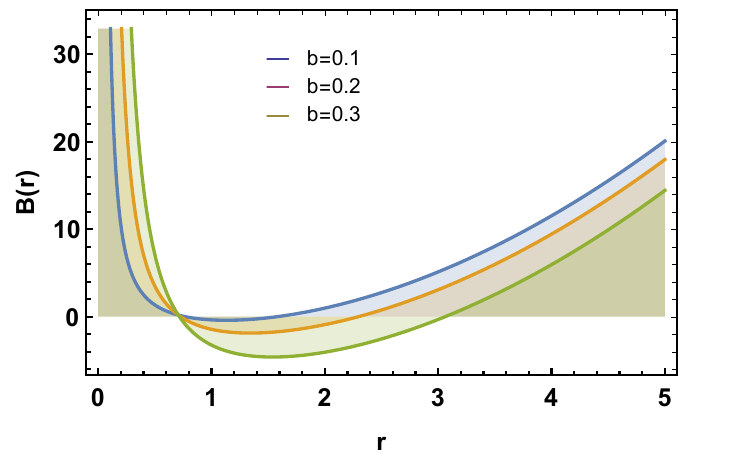}\\
(a)\\
\includegraphics[height=5cm,width=7cm]{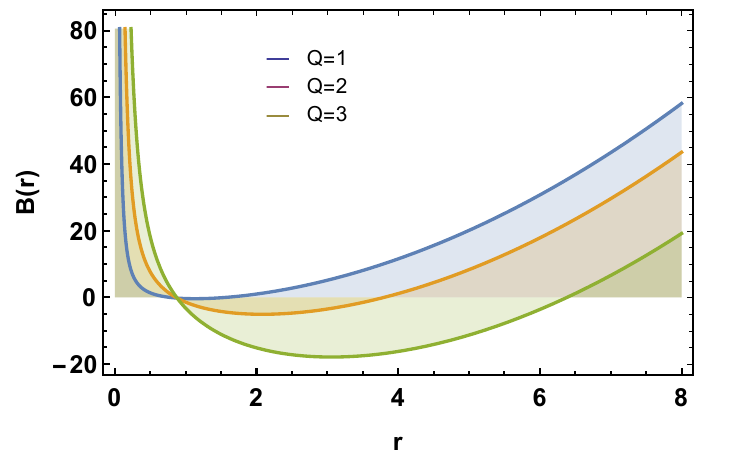}
\includegraphics[height=5cm,width=7cm]{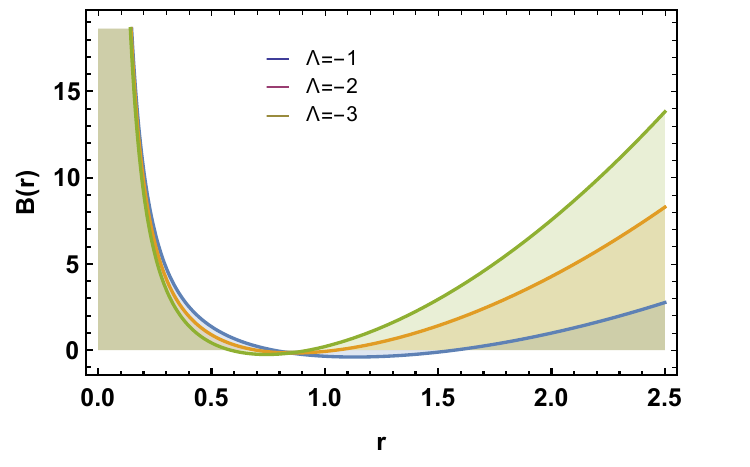}\\
(a) \hspace{8cm} (b)
\caption{Metric coefficient $B(r)$.(a) $\Lambda=-1$ and $Q=M=r_0=1$. (b) $\Lambda=-1$, $b=0.1$ and $M=r_0=1$. (c) $b=0.1$ and $Q=M=r_0=1$.}
\label{fig3}
\end{figure}

Figure (\ref{fig2}) illustrates the behavior of the metric coefficient $A(r)$, which corresponds to the temporal component of the charged BTZ-type black hole metric modified by Bopp–Podolsky electrodynamics. In panel (a), $A(r)$ is plotted for different values of the Bopp–Podolsky parameter $b$, with fixed parameters $\Lambda = -1$ and $Q = M = r_0 = 1$. The curves demonstrate that as $b$ increases, the deformation in the metric becomes more pronounced near the origin, indicating the influence of higher-derivative electrodynamic effects. Panel (b) shows how $A(r)$ varies with different values of the electric charge $Q$, keeping $\Lambda = -1$, $b = 0.1$, and $M = r_0 = 1$. As $Q$ increases, the metric coefficient exhibits a sharper deviation, reflecting the stronger electromagnetic contribution to the geometry. Also, in panel (c), the effect of the cosmological constant $\Lambda$ is examined for values $\Lambda = -1, -2, -3$, with fixed $b = 0.1$ and $Q = M = r_0 = 1$. More negative values of $\Lambda$ deepen the gravitational potential well, which is visible as a stronger curvature in $A(r)$.

Figure (\ref{fig3}) shows the behavior of the metric coefficient $B(r)$, which appears in the radial component of the BTZ-type black hole line element under Bopp–Podolsky electrodynamics. In panel (a), the evolution of $B(r)$ is presented for various values of the Bopp–Podolsky parameter $b$, with fixed $\Lambda = -1$ and $Q = M = r_0 = 1$. The results indicate that increasing $b$ causes greater deviations in the radial metric function near the origin, highlighting the role of higher-order curvature corrections. Panel (b) explores how $B(r)$ changes with different electric charge values $Q$, keeping $\Lambda = -1$, $b = 0.1$, and $M = r_0 = 1$ constant. In this context, a larger $Q$ parameter of the electromagnetic contribution leads to more substantial alterations in the metric profile. Also, panel (c) investigates the influence of the cosmological constant $\Lambda$, taking values $-1, -2, -3 $ with fixed $b = 0.1$ and $Q = M = r_0 = 1$. As $|\Lambda|$ increases, the radial function exhibits a steeper descent, indicating the deepening of the effective gravitational potential.

\begin{figure}[ht]
\centering
\includegraphics[height=5cm,width=7cm]{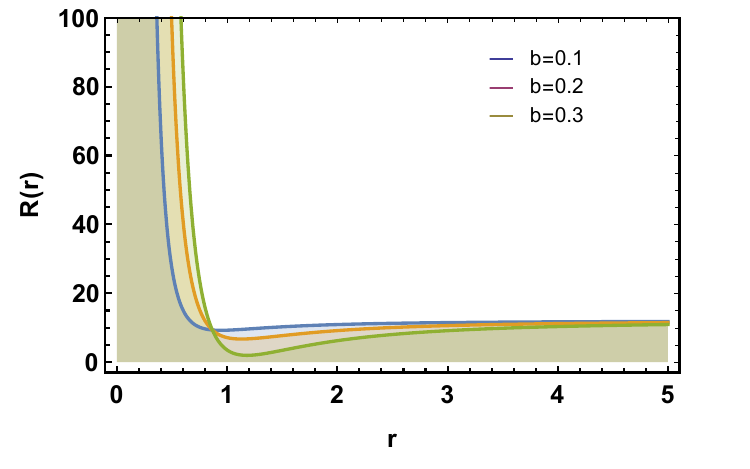}\\
(a)\\
\includegraphics[height=5cm,width=7cm]{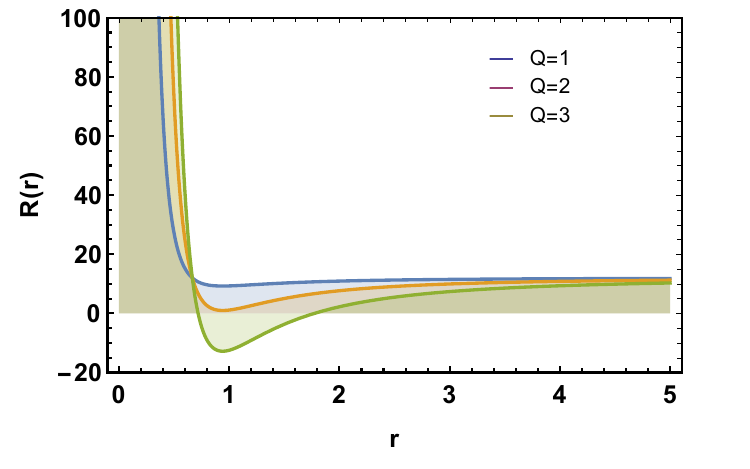}
\includegraphics[height=5cm,width=7cm]{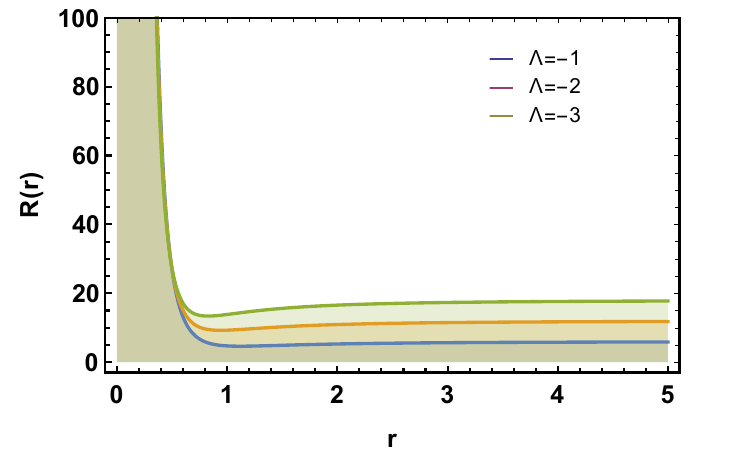}\\
(b) \hspace{8cm} (c)
\caption{Curvature scalar $R$.(a) $\Lambda=-1$ and $Q=M=r_0=1$. (b) $\Lambda=-1$, $b=0.1$ and $M=r_0=1$. (c) $b=0.1$ and $Q=M=r_0=1$.}
\label{fig2.1}
\end{figure}

\begin{figure}[ht]
\centering
\includegraphics[height=5cm,width=7cm]{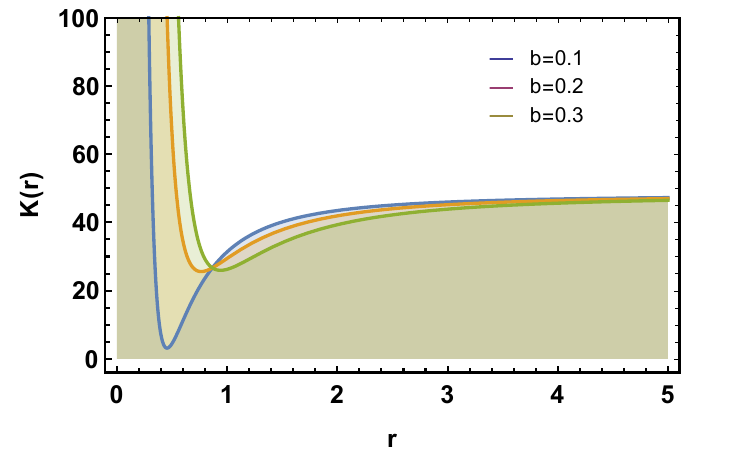}\\
(a)\\
\includegraphics[height=5cm,width=7cm]{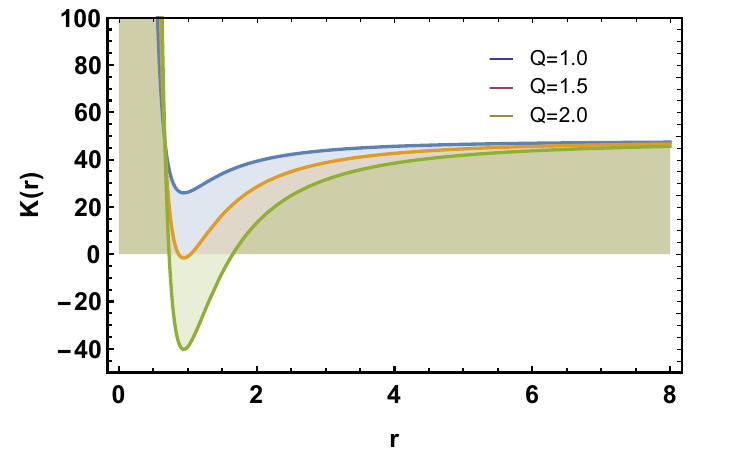}
\includegraphics[height=5cm,width=7cm]{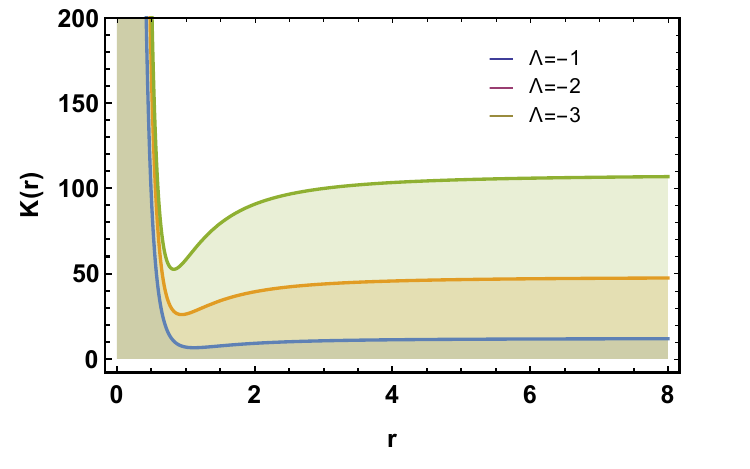}\\
(a) \hspace{8cm} (b)
\caption{Kretschmann scalar $K$.(a) $\Lambda=-1$ and $Q=M=r_0=1$. (b) $\Lambda=-1$, $b=0.1$ and $M=r_0=1$. (c) $b=0.1$ and $Q=M=r_0=1$.}
\label{fig3.1}
\end{figure}

{Figures (\ref{fig2.1}) and (\ref{fig3.1}) illustrate the behavior of the curvature scalar $R(r)$ and the Kretschmann scalar $K(r)$, respectively, for the charged BTZ-type BH in the context of Bopp–Podolsky electrodynamics, show the influence of the Bopp–Podolsky parameter $b$, the electric charge $Q$, and the cosmological constant $\Lambda$. In Figure (\ref{fig2.1})(a), where $\Lambda = -1$ and $Q = M = r_0 = 1$, the scalar curvature $R(r)$ is plotted for three values of the Bopp–Podolsky parameter $b = 0.1, 0.2, 0.3$, as $b$ increases, the curvature grows more sharply near the origin, reflecting the stronger impact of higher-derivative terms. In Figure (\ref{fig2.1})(b), with $\Lambda = -1$, $b = 0.1$, and $M = r_0 = 1$, the variation of $R(r)$ with electric charge $Q = 1, 2, 3$ shows that increasing $Q$ intensifies the curvature near the core. Figure 4(c) displays the effect of the cosmological constant $\Lambda = -1, -2, -3$, using fixed $b = 0.1$ and $Q = M = r_0 = 1$, more negative values of $\Lambda$ increase the curvature magnitude, indicating stronger spacetime warping in an AdS background. Similarly, Figure (\ref{fig3.1}) shows the Kretschmann scalar $K(r)$, which captures the square of the Riemann tensor and provides a more complete measure of curvature singularities. In panel (\ref{fig3.1})(a), for $\Lambda = -1$ and $Q = M = r_0 = 1$, the scalar is plotted against $r$ for $b = 0.1, 0.2, 0.3$, and higher values of $b$ lead to sharper divergences near $r = 0$, confirming that the curvature intensifies with stronger higher-derivative effects. In Figure (\ref{fig3.1})(b), with $\Lambda = -1$, $b = 0.1$, and $M = r_0 = 1$, the Kretschmann scalar is shown for charges $Q = 1, 2, 3$, as expected, larger $Q$ values produce larger curvature peaks, due to enhanced electromagnetic energy density. Also, Figure (\ref{fig3.1})(c) tested $\Lambda = -1, -2, -3$ with $b = 0.1$ and $Q = M = r_0 = 1$, the results reveal that increasing the magnitude of the negative cosmological constant amplifies the spacetime curvature significantly, particularly close to the BH core, further underscoring the role of $\Lambda$ in determining the gravitational environment under Bopp–Podolsky corrections.}


%%%%%%%%%%%%%%%%%%%%%%%%%%%%%%%%%%%%%%%%%%%%%%%%%%%%%%%%%%%%%%%%%%%%%%%%

\begin{figure}[ht]
\centering
\includegraphics[height=5cm,width=7cm]{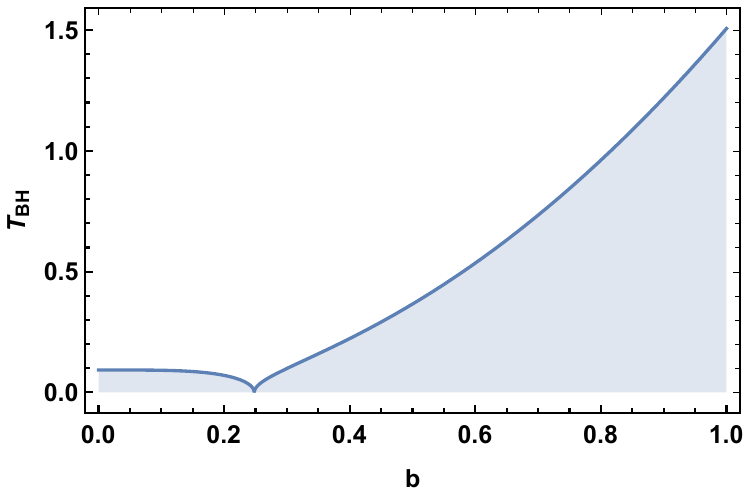}\\
(a)\\
\includegraphics[height=5cm,width=7cm]{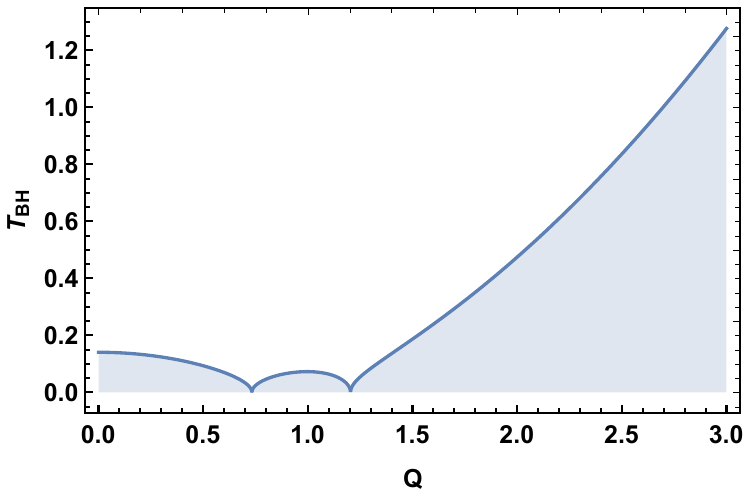}
\includegraphics[height=5cm,width=7cm]{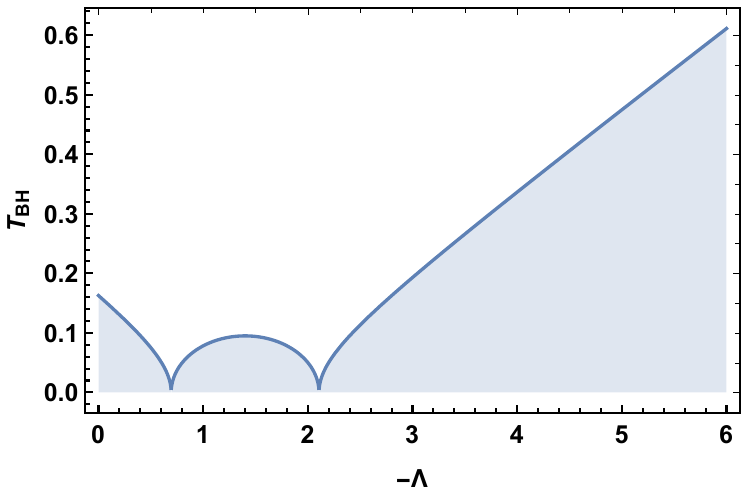}\\
(b) \hspace{8cm} (c)
\caption{Hawking Temperature. (a) $\Lambda=-1$ and $Q=M=r_0=1$. (b) $\Lambda=-1$, $b=0.1$ and $M=r_0=1$. (c) $b=0.1$ and $Q=M=r_0=1$.}
\label{fig4}
\end{figure}

Figure (\ref{fig4}) illustrate the behavior of the Hawking temperature $T_{BH}$ of the charged BTZ-type BH,  in panel (a), $T_{BH}$ is plotted as a function of the Bopp–Podolsky parameter $b$, with fixed values $\Lambda = -1$ and $Q = M = r_0 = 1$. The plot reveals that the Hawking temperature increases monotonically with $b$, indicating that the higher-derivative corrections enhance the surface gravity of the black hole. Panel (b) shows the temperature as a function of the electric charge $Q$, for fixed $\Lambda = -1$, $b = 0.1$, and $M = r_0 = 1$. Also, panel (c) illustrates the variation of $T_{BH}$ with under the cosmological constant $\Lambda$ parameters, using values $\Lambda = -1, -2, -3$ and fixed $b = 0.1$, $Q = M = r_0 = 1$. As $|\Lambda|$ increases, the Hawking temperature decreases, indicating a cooling effect due to the stronger negative curvature of the spacetime.

\subsection{Second Approach}

Now, considering as a solution the second order approximation of Eq.(\ref{PertE}) and linearizing around $\xi = 0$ and considering terms up to order $\xi^2$, we have
\begin{align}
 \frac{1}{2r}B_{2}'(r)&+\frac{Q^{2}}{r^{2}(M+\Lambda r^{2})}\left(A_{2}(r)-B_{2}(r)\right)+\frac{2Q}{r}E_{2}(r)
 \nonumber\\ &=\frac{4\mathit{b}^{2}Q^{2}}{\xi^{2}r^{6}\left(M+\Lambda r^{2}\right)}\left[2MQ^{2}r^{2}+4\mathit{b}^{2}\left(Q^{2}+M\right)\left(Q^{2}M-\Lambda r^{2}\left(M-3Q^{2}\right)\right)\right.\nonumber\\&\left.+\Lambda r^{4}\left(2Q^{2}+\mathit{b}^{2}\Lambda\left(-M+4Q^{2}+3\Lambda r^{2}\right)\right)-2Q^{2}r^{2}\left(M+2Q^{2}+\Lambda r^{2}\right)\ln\left(\frac{r}{r_{0}}\right)\right],
 \nonumber\\
 \frac{1}{2r}A_{2}'(r)&+\frac{(Q^{2}-\Lambda r^{2})}{r^{2}(M+\Lambda r^{2})}\left(A_{2}(r)-B_{2}(r)\right)+\frac{2Q}{r}E_{2}(r)
  \nonumber\\ &=\frac{4\mathit{b}^{2}Q^{2}}{\xi^{2}r^{6}\left(M+\Lambda r^{2}\right)}\left[4\mathit{b}^{2}\left(\Lambda r^{2}\left(M^{2}+3Q^{4}\right)+MQ^{2}\left(M+Q^{2}\right)\right)\right.\nonumber\\&\left.+\Lambda r^{4}\left(\mathit{b}^{2}\Lambda\left(15M-12Q^{2}+11\Lambda r^{2}\right)+6Q^{2}\ln\left(\frac{r}{r_{0}}\right)\right)-2Q^{2}r^{2}\left(2Q^{2}-M\right)\ln\left(\frac{r}{r_{0}}\right)\right], 
 \nonumber\\
 \frac{1}{2}A_{2}''(r)&-\frac{\Lambda r}{2(M+\Lambda r^{2})}\left(A_{2}'(r)-B_{2}'(r)\right)-\frac{\big(MQ^{2}+(M+Q^{2})\Lambda r^{2}\big)}{r^{2}(M+\Lambda r^{2})^{2}}\left(A_{2}(r)-B_{2}(r)\right)-\frac{2Q}{r}E_{2}(r) 
 \nonumber \\& -\frac{4\mathit{b}^{2}Q^{2}}{\xi^{2}r^{6}\left(M+\Lambda r^{2}\right)}\left\{2M\left[2\mathit{b}^{2}\left(\Lambda r^{2}\left(3M^{2}+9MQ^{2}+4Q^{4}\right)+MQ^{2}\left(5M+Q^{2}\right)\right)-3MQ^{2}r^{2}\right]\right.\nonumber\\&+\Lambda r^{4}\left(3\mathit{b}^{2}\Lambda\left(9M^{2}+M\left(4Q^{2}+6\Lambda r^{2}\right)+\left(\Lambda r^{2}-2Q^{2}\right)^{2}\right)-2Q^{2}\left(7M+4\Lambda r^{2}\right)\right)\nonumber\\&\left.+2Q^{2}r^{2}\left(3M^{2}-2M\left(Q^{2}-3\Lambda r^{2}\right)-\Lambda r^{2}\left(2Q^{2}-5\Lambda r^{2}\right)\right)\ln\left(\frac{r}{r_{0}}\right)\right\}.  
\label{Equations32}
\end{align}
As in the previous analysis, all terms involving products of the type $\xi^2 a^2$, $\xi^2 b^2$ or cubic (or higher) powers of $\xi$ are eliminated, since they contribute only at higher orders of the perturbative expansion. The exclusion of terms such as $a^2\xi^2$, $b^2\xi^2$ and $\xi^3$ is reiterated, as well as any higher order terms. Furthermore, as in the case discussed previously, there is no effective contribution from $a^2$, since this parameter only emerges at higher perturbative orders. 
With this, it is possible to express the line element and the electric field corresponding to the Einstein-Bopp-Podolsky system, including corrections up to the second order in the perturbative expansion, as follows
\begin{align}
    ds^{2}=&\Big[-M-\Lambda r^{2}+\Sigma_2(r)\Big]dt^{2}-\frac{dr^{2}}{\Big[-M-\Lambda r^{2}+\Omega_2(r)\Big]}
	-r^{2}d\theta^{2},\nonumber\\
    E_2(r)=&\frac{Q}{r}\left(1+2b^{2}\Lambda+6b^{4}\Lambda^{2}\right)+\frac{8b^{2}Q^{3}}{r^{3}}\left(1+b^{2}\Lambda\right)+\frac{8b^{4}Q^{5}}{r^{5}},
\end{align}
where
 \begin{align}
\Sigma_2(r)=&-\frac{4b^{2}Q^{2}}{r^{2}}\left(M+4b^{2}M\Lambda-Q^{2}\right)-\frac{8b^{4}Q^{4}M}{r^{4}}-\left(2Q^{2}+\frac{8b^{2}Q^{4}}{r^{2}}\right)\ln\left(\frac{r}{r_{0}}\right),\nonumber \\
 \Omega_2(r)=&8b^{2}Q^{2}\Lambda\left(1+4b^{2}\Lambda\right)+\frac{4b^{2}Q^{2}}{r^{2}}\left(M+Q^{2}+4b^{2}\Lambda (M-2Q^2)\right)-\frac{8b^{4}Q^{4}M}{r^{4}}\nonumber \\ &-\left(2Q^{2}-\frac{8b^{2}Q^{4}}{r^{2}}\right)\ln\left(\frac{r}{r_{0}}\right).
 \end{align}
From this expression, it is observed that all corrections associated with the parameter $b$ present a rapid decay in the asymptotic region, which indicates that the zero-order solution of the BTZ type is modified by terms that behave as negative powers of the radial coordinate. This behavior is also manifested in the logarithmic term, which also contributes corrections that become negligible at large distances.

\begin{figure}[ht]
\centering
\includegraphics[height=5cm,width=7cm]{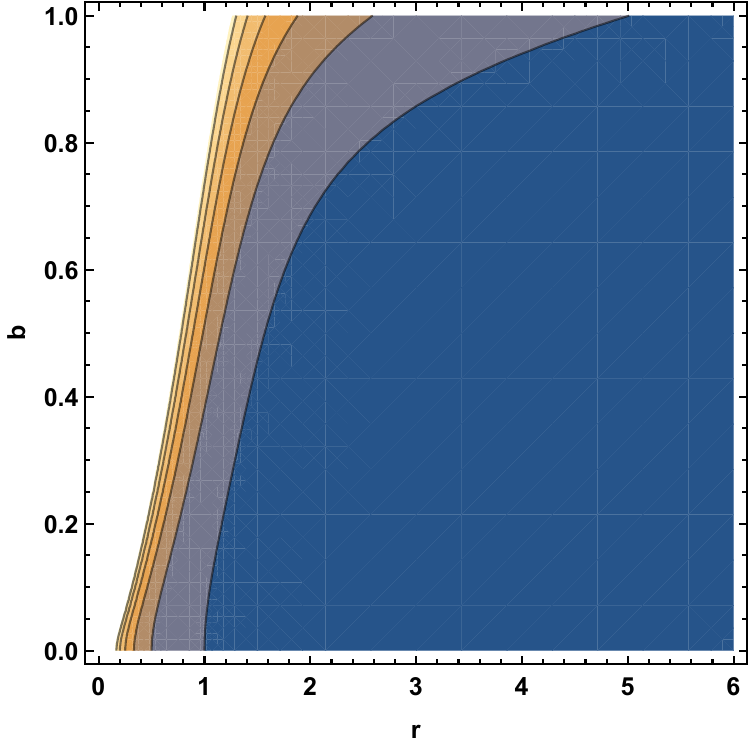}\\
(a)\\
\includegraphics[height=5cm,width=7cm]{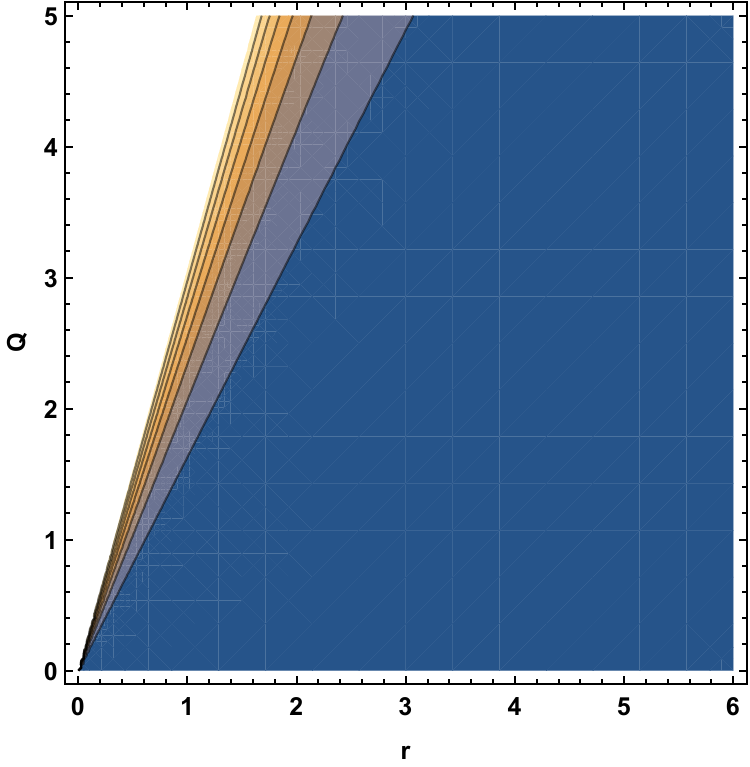}
\includegraphics[height=5cm,width=7cm]{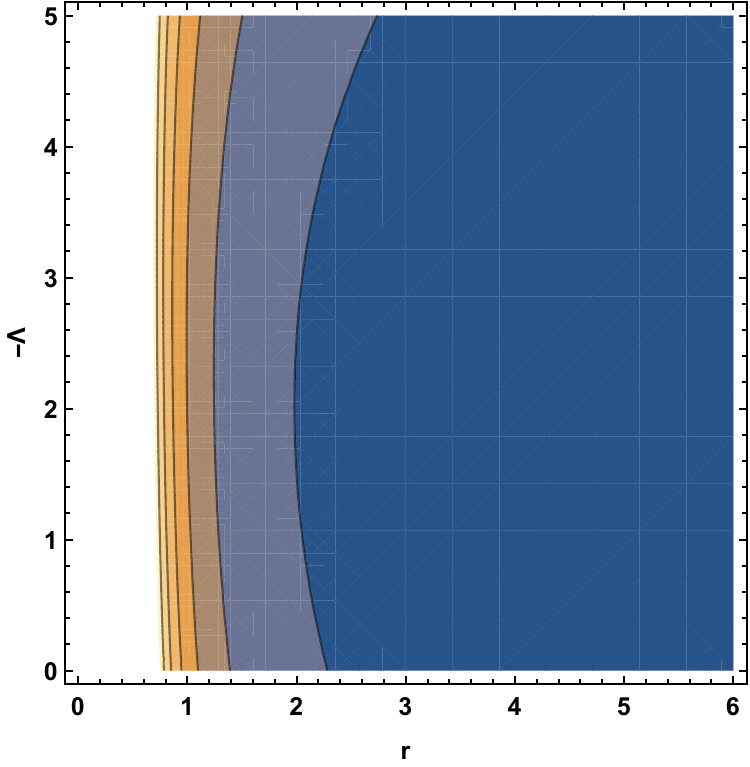}\\
(b) \hspace{8cm} (c)
\caption{Electric field. (a) $\Lambda=-1$ and $Q=M=r_0=1$. (b) $\Lambda=-1$, $b=0.1$ and $M=r_0=1$. (c) $b=0.1$ and $Q=M=r_0=1$.}
\label{fig5}
\end{figure}
\begin{figure}[ht]
\centering
\includegraphics[height=5cm,width=7cm]{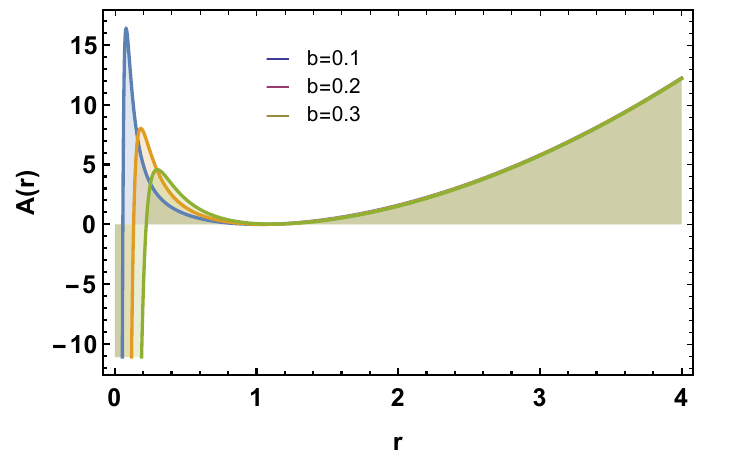}\\
(a)\\
\includegraphics[height=5cm,width=7cm]{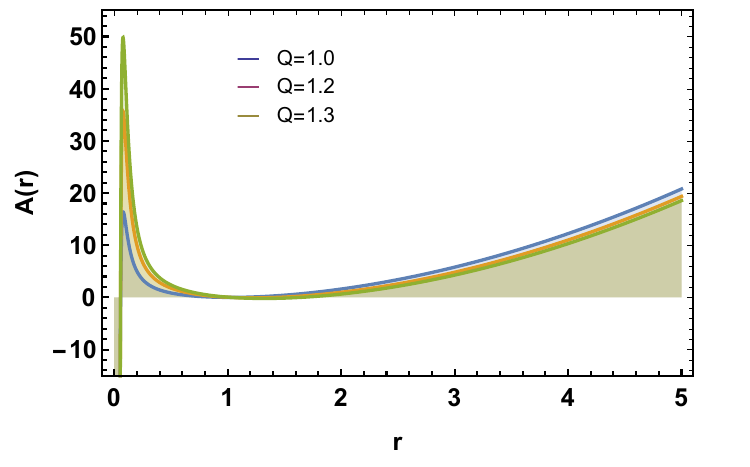}
\includegraphics[height=5cm,width=7cm]{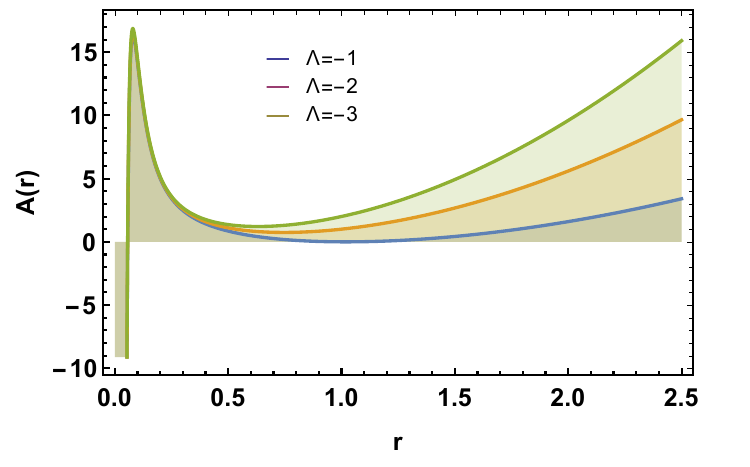}\\
(b) \hspace{8cm} (c)
\caption{Metric coefficient $A(r)$.(a) $\Lambda=-1$ and $Q=M=r_0=1$. (b) $\Lambda=-1$, $b=0.1$ and $M=r_0=1$. (c) $b=0.1$ and $Q=M=r_0=1$.}
\label{fig6}
\end{figure}
\begin{figure}[ht]
\centering
\includegraphics[height=5cm,width=7cm]{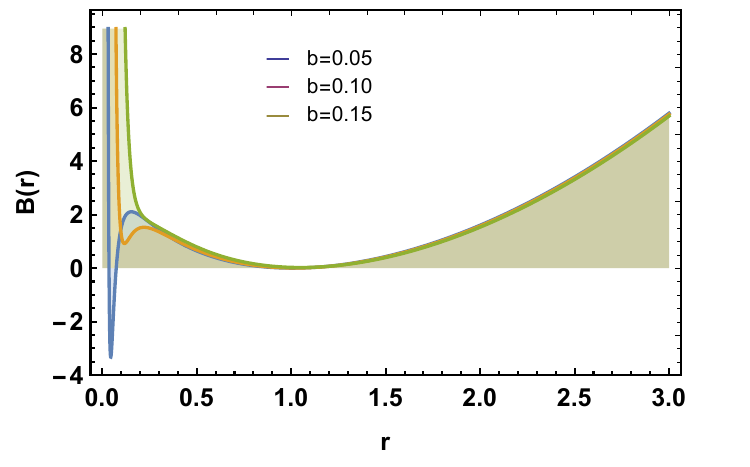}\\
(a)\\
\includegraphics[height=5cm,width=7cm]{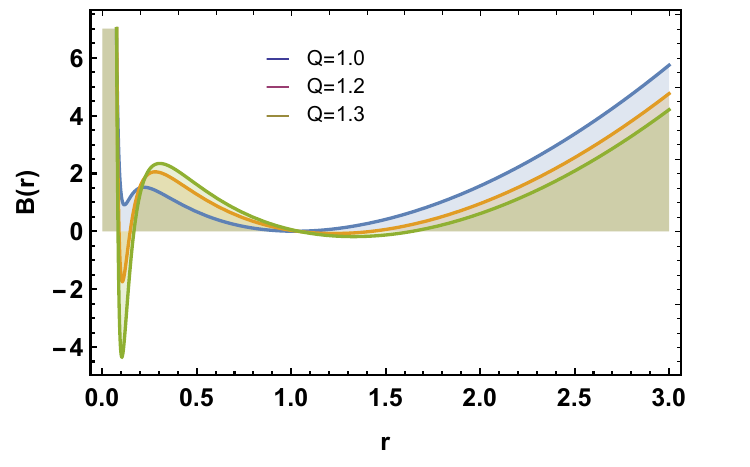}
\includegraphics[height=5cm,width=7cm]{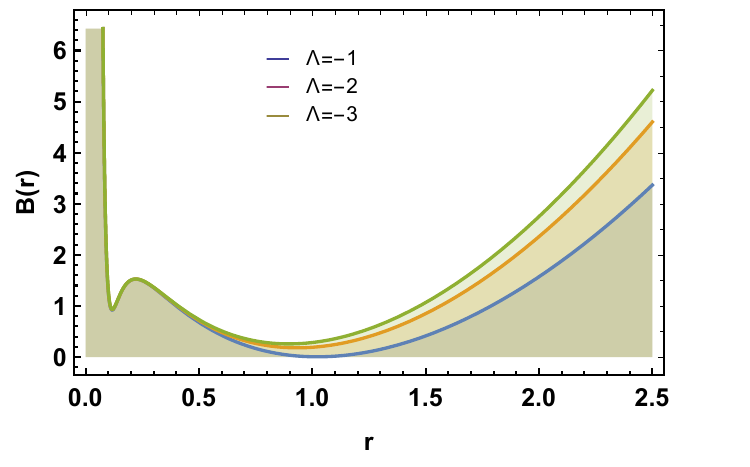}\\
(b) \hspace{8cm} (c)
\caption{Metric coefficient $B(r)$.(a) $\Lambda=-1$ and $Q=M=r_0=1$. (b) $\Lambda=-1$, $b=0.1$ and $M=r_0=1$. (c) $b=0.1$ and $Q=M=r_0=1$.}
\label{fig7}
\end{figure}

\begin{figure}[ht]
\centering
\includegraphics[height=5cm,width=7cm]{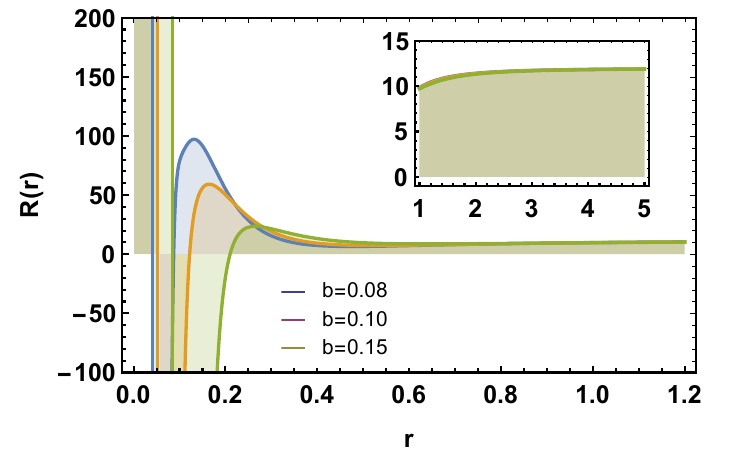}\\
(a)\\
\includegraphics[height=5cm,width=7cm]{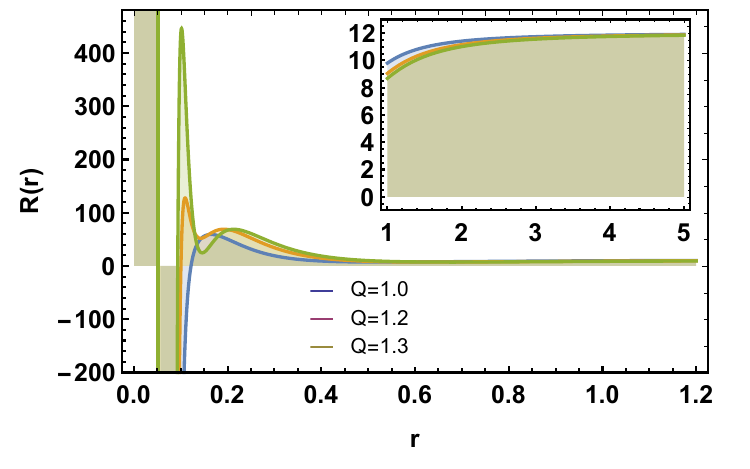}
\includegraphics[height=5cm,width=7cm]{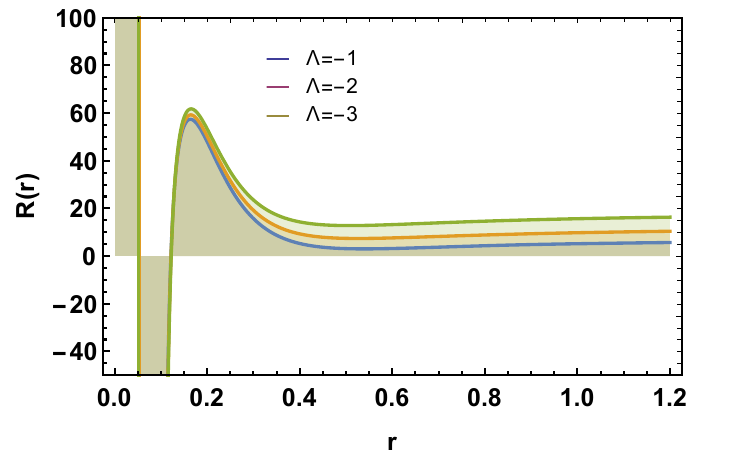}\\
(b) \hspace{8cm} (c)
\caption{Curvature scalar $R$. (a) $\Lambda=-1$ and $Q=M=r_0=1$. (b) $\Lambda=-1$, $b=0.1$ and $M=r_0=1$. (c) $b=0.1$ and $Q=M=r_0=1$.}
\label{fig6.1}
\end{figure}

\begin{figure}[ht]
\centering
\includegraphics[height=5cm,width=7cm]{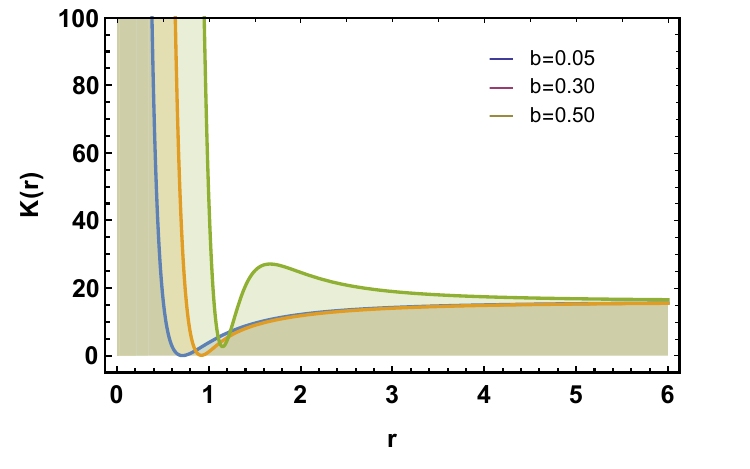}\\
(a)\\
\includegraphics[height=5cm,width=7cm]{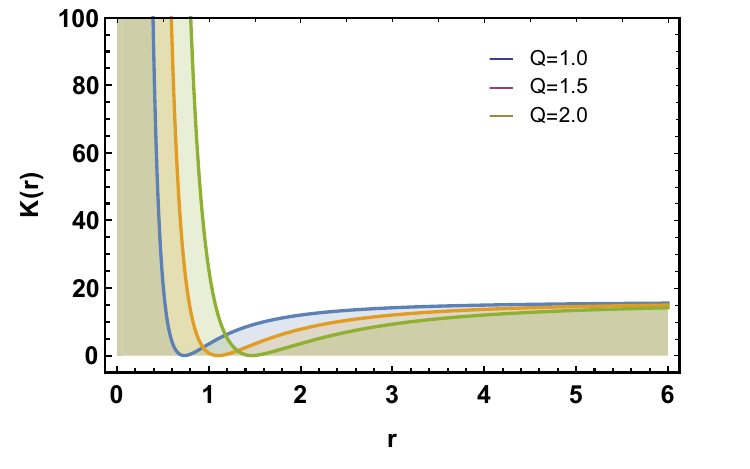}
\includegraphics[height=5cm,width=7cm]{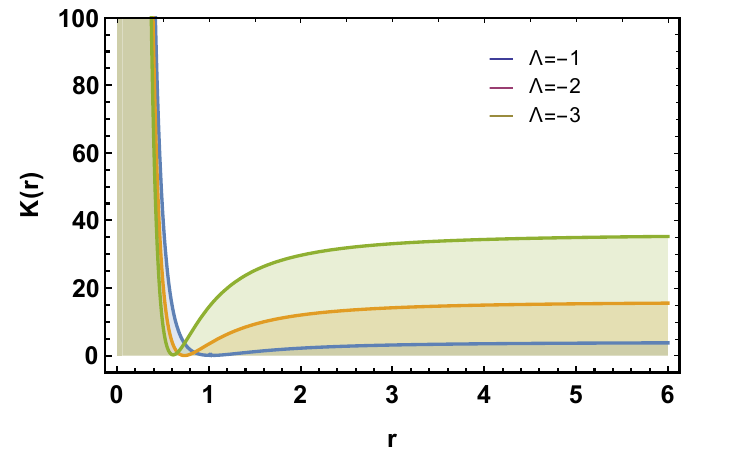}\\
(b) \hspace{8cm} (c)
\caption{Kretschmann scalar $K$. (a) $\Lambda=-1$ and $Q=M=r_0=1$. (b) $\Lambda=-1$, $b=0.1$ and $M=r_0=1$. (c) $b=0.1$ and $Q=M=r_0=1$.}
\label{fig7.1}
\end{figure}

{Figures (\ref{fig6.1}) and (\ref{fig7.1}) illustrate detailed analyses of the curvature scalar $R(r)$ and the Kretschmann scalar $K(r)$, respectively, under the influence of Bopp–Podolsky electrodynamics for a charged BTZ-type BH, with explicit focus on how the Bopp–Podolsky parameter $b$, electric charge $Q$, and cosmological constant $\Lambda$ affect the spacetime geometry. In Figure (\ref{fig6.1})(a), the scalar curvature $R(r)$ is plotted against the radial coordinate $r$ for values $b = 0.08, 0.10, 0.15$ while fixing $\Lambda = -1$ and $Q = M = r_0 = 1$, as $b$ increases, the curvature becomes significantly more divergent near the origin, indicating an enhancement of the gravitational effects due to the higher-derivative electrodynamics. Figure (\ref{fig6.1})(b) explores $R(r)$ as a function of $r$ for different electric charges $Q = 1.0, 1.2, 1.3$, with $\Lambda = -1$, $b = 0.1$, and $M = r_0 = 1$, increasing $Q$ results in more pronounced curvature peaks, especially near the horizon, reflecting the stronger electromagnetic contribution to the geometry. In Figure (\ref{fig6.1})(c), the scalar curvature is shown for cosmological constants $\Lambda = -1, -2, -3$, keeping $b = 0.1$ and $Q = M = r_0 = 1$ fixed, as the magnitude of $\Lambda$ increases, the spacetime exhibits more intense curvature across all radial values, signifying the deepening of the AdS gravitational potential due to the negative cosmological constant. Also, Figure (\ref{fig7.1}) depicts the Kretschmann scalar $K(r)$, which encapsulates the square of the Riemann tensor and provides a more comprehensive measure of spacetime curvature. In this case, in Figure (\ref{fig7.1})(a), $K(r)$ is plotted for $b = 0.05, 0.30, 0.50$ with fixed $\Lambda = -1$ and $Q = M = r_0 = 1$, larger values of $b$ dramatically amplify the curvature near $r = 0$, illustrate the impact of higher-derivative terms at short distances. Figure (\ref{fig7.1})(b) shows $K(r)$ for electric charges $Q = 1.0, 1.5, 2.0$, using $\Lambda = -1$, $b = 0.1$, and $M = r_0 = 1$, increasing $Q$ intensifies the curvature profile, particularly in the near-horizon region, due to the rising energy density from the electric field. Finally, Figure (\ref{fig7.1})(c) explores the effect of the cosmological constant by varying $\Lambda = -1, -2, -3$ with $b = 0.1$ and $Q = M = r_0 = 1$, as $|\Lambda|$ increases, the Kretschmann scalar shows a significant increase in curvature throughout the domain, further emphasizing the role of AdS geometry in enhancing gravitational strength in the Bopp–Podolsky-corrected spacetime.}


%%%%%%%%%%%%%%%%%%%%%%%%%%%%%%%%%%%%%%%%%%%%%%%%%%%%%%%%%%%%%%%%%%%%%%%%%%%%%%%%%%%%%%%%%%%

\begin{figure}[ht]
\centering
\includegraphics[height=5cm,width=7cm]{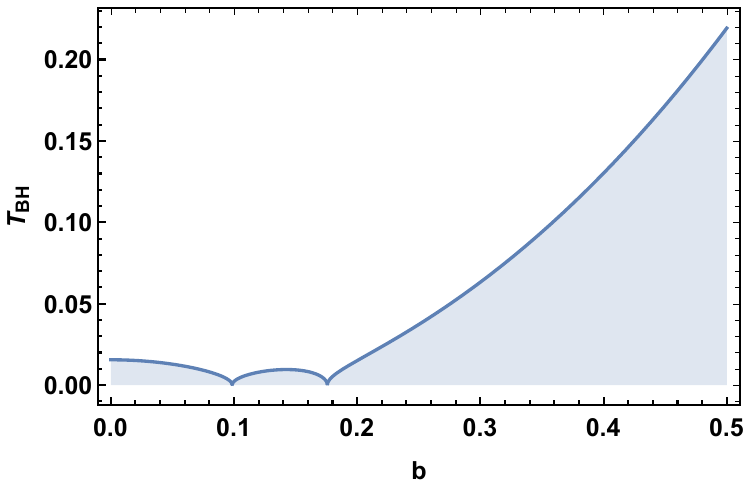}\\
(a)\\
\includegraphics[height=5cm,width=7cm]{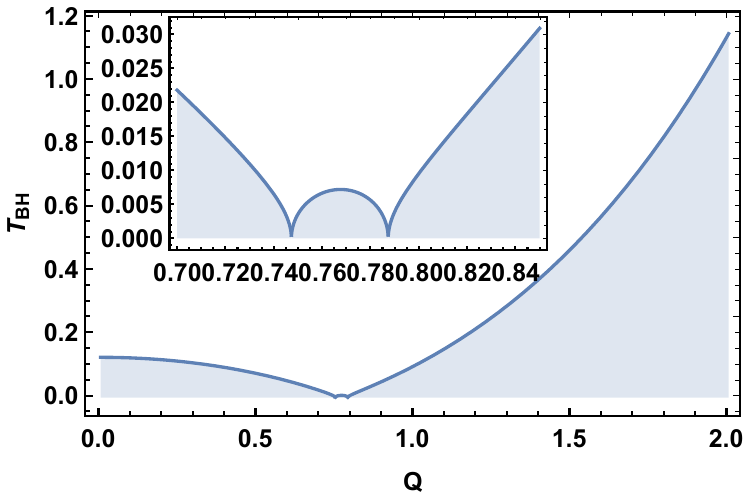}
\includegraphics[height=5cm,width=7cm]{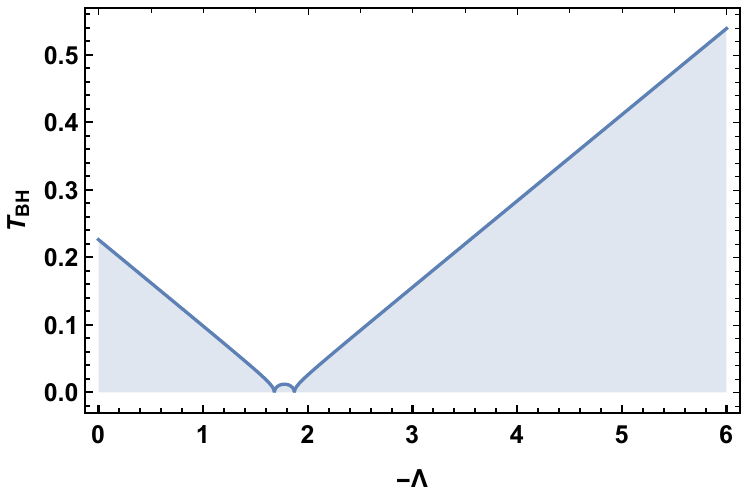}\\
(b) \hspace{8cm} (c)
\caption{Hawking Temperature. (a) $\Lambda=-1$ and $Q=M=r_0=1$. (b) $\Lambda=-1$, $b=0.1$ and $M=r_0=1$. (c) $b=0.1$ and $Q=M=r_0=1$.}
\label{fig8}
\end{figure}
Figure (\ref{fig5}) provides a second set of visualizations for the electric field $E(r)$ in the charged BTZ-type black hole solution influenced by Bopp–Podolsky electrodynamics, reaffirming the effects of key parameters. In panel (a), the electric field is shown as a function of the radial coordinate $r$ for different values of the Bopp–Podolsky parameter $b$, with fixed $\Lambda = -1$ and $Q = M = r_0 = 1$. The field strength increases with $b$, especially near the origin, due to the enhanced contribution of higher-order derivative terms in the theory. Panel (b) examines how the electric field responds to changes in the electric charge $Q$, for fixed $\Lambda = -1$, $b = 0.1$, and $M = r_0 = 1$. Also, increasing $Q$ amplifies the magnitude of the electric field across the entire range of $r$. In panel (c), the influence of the cosmological constant $\Lambda$ is investigated by varying it while keeping $b = 0.1$ and $Q = M = r_0 = 1$. Larger values of $|\Lambda|$ slightly shift the electric field profile, showing the effect of $b$ and $Q$ parameters.

Figure (\ref{fig6}) illustrates the behavior of the metric coefficient $A(r)$, corresponding to the temporal component of the black hole metric, incorporating second-order corrections from Bopp–Podolsky electrodynamics. In panel (a), the influence of the Bopp–Podolsky parameter $b$ is analyzed for $b = 0.1, 0.2, 0.3$, with fixed values $\Lambda = -1$ and $Q = M = r_0 = 1$. The results show that increasing $b$ introduces stronger deviations in $A(r)$ near the origin, confirming the effect of higher-derivative terms on the geometry. Panel (b) teste the dependence of $A(r)$ on the electric charge $Q = 1.0, 1.2, 1.3$, with parameters $\Lambda = -1$, $b = 0.1$, and $M = r_0 = 1$ held constant. The plot indicates that larger charge values significantly modify the metric function, producing deeper gravitational wells. In panel (c), the effect of the cosmological constant is presented for $\Lambda = -1, -2, -3$, while $b = 0.1$ and $Q = M = r_0 = 1$ remain fixed. As $|\Lambda|$ increases.

Figure (\ref{fig7}) shows the behavior of the metric coefficient $B(r)$, associated with the radial component of the black hole metric, incorporating second-order perturbative corrections from Bopp–Podolsky electrodynamics. In panel (a), the function $B(r)$ is plotted for varying Bopp–Podolsky parameters $b = 0.05, 0.10, 0.15$, with fixed values $\Lambda = -1$ and $Q = M = r_0 = 1$. The curves indicate that as $b$ increases, the radial component of the metric deviates more significantly from the classical BTZ profile, especially near the origin. Panel (b) explores the dependence of $B(r)$ on the electric charge $Q = 1.0, 1.2, 1.3$, while keeping $\Lambda = -1$, $b = 0.1$, and $M = r_0 = 1$ constant. The results show that increasing $Q$ leads to a stronger deformation in the geometry, consistent with a higher electromagnetic contribution. In panel (c), the effect of the cosmological constant is illustrated for $\Lambda = -1, -2, -3$, with fixed $b = 0.1$ and $Q = M = r_0 = 1$. The plots demonstrate that a more negative $\Lambda$ results in a steeper and more rapidly changing $B(r)$, indicating stronger curvature in the AdS background. Together, these results emphasize the sensitivity of the radial metric component to Bopp–Podolsky corrections, charge, and spacetime curvature.
 
Figure~(\ref{fig8}) illustrates the behavior of the Hawking temperature $ T_{H} $ about various parameters of a black hole system. In panel (a), the temperature is plotted against the Bopp-Podolsky parameter $ b $ while keeping the cosmological constant $\Lambda = -1$ and charge $ Q = M = r_0 = 1 $ constant. For $ b $ increases, $ T_H $ initially decreases to a minimum before rising again, indicating a non-linear response to changes in the perturbative parameter. Panel (b) examines the dependence of $ T_H $ on the electric charge $ Q $, with fixed values $\Lambda = -1$, $ b = 0.1 $, and $ M = r_0 = 1 $. Also, the results show a gradual increase in temperature with $ Q $, suggesting that higher charge contributes to a greater Hawking temperature, which is further emphasized by the inset that shows a local minimum. In this context, panel (c) tests the effect of varying the cosmological constant $-\Lambda$ for fixed $ b = 0.1 $ and $ Q = M = r_0 = 1 $. The plot indicates that as $-\Lambda$ increases, $ T_H $ decreases, showing the significant influence of spacetime curvature on black hole models.

\section{Thermodynamic properties}

For a more complete analysis, in this section we study thermodynamic properties, especially entropy and heat capacity. Using the temperature of the black hole, we can define entropy as a function of the event horizon radius $r_H$, following the prescription
\begin{eqnarray}
S=\int \frac{dM}{T_{BH}}.
\end{eqnarray}
With the entropy at hand, we can directly compute the heat capacity $C$, where
\begin{eqnarray}
C = T\, \frac{\partial S}{\partial T}.
\end{eqnarray}
Let is now apply them to our first and second approximation solutions.

\subsection{First Approach}

\begin{figure}[ht]
\centering
\includegraphics[height=5cm,width=7cm]{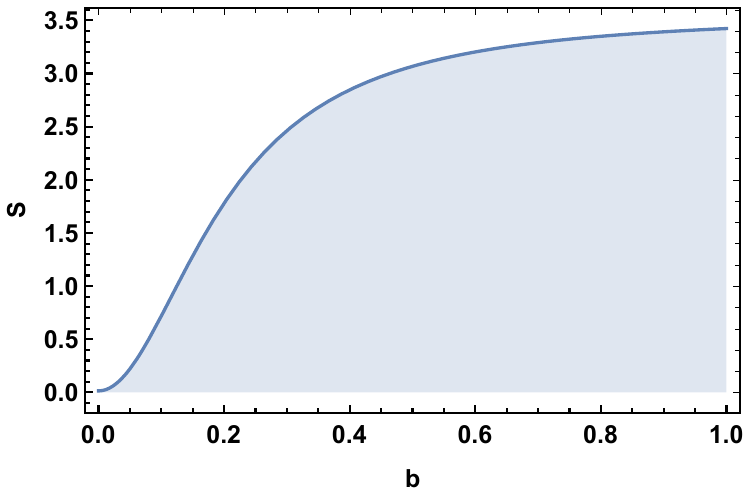}\\
(a)\\
\includegraphics[height=5cm,width=7cm]{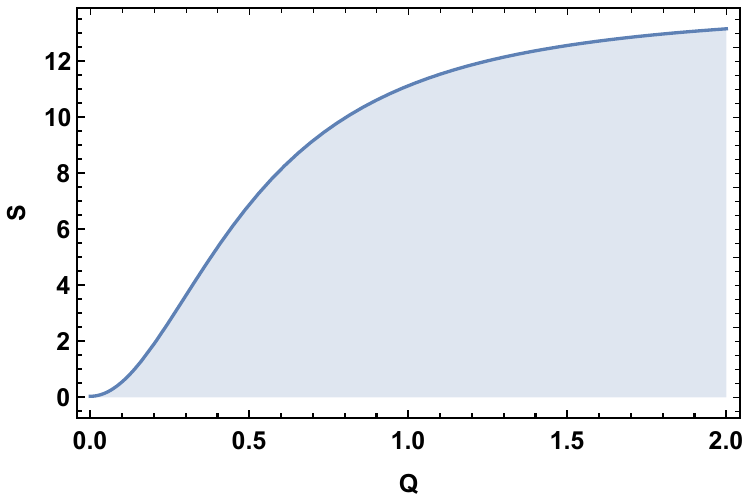}
\includegraphics[height=5cm,width=7cm]{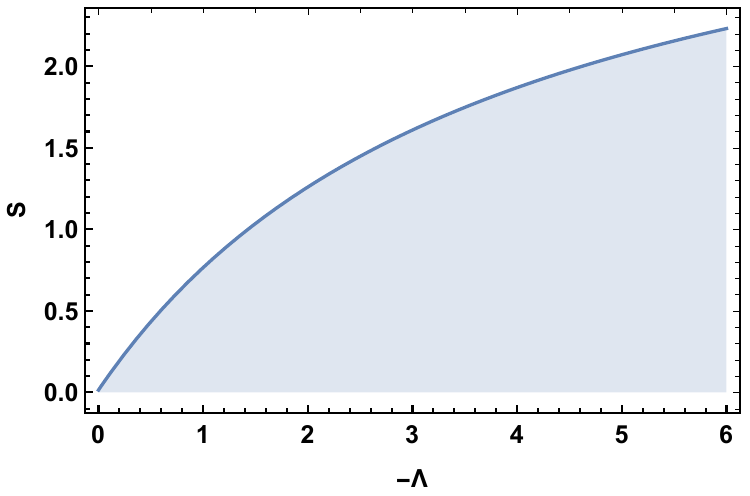}\\
(b) \hspace{8cm} (c)
\caption{Entropy. (a) $\Lambda=-1$ and $Q=M=r_0=1$. (b) $\Lambda=-1$, $b=0.1$ and $M=r_0=1$. (c) $b=0.1$ and $Q=M=r_0=1$.}
\label{fig13}
\end{figure}

\begin{figure}[ht]
\centering
\includegraphics[height=5cm,width=7cm]{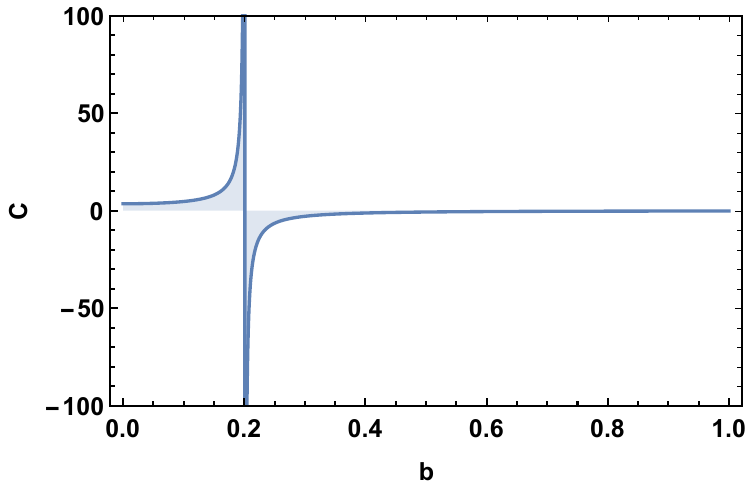}\\
(a)\\
\includegraphics[height=5cm,width=7cm]{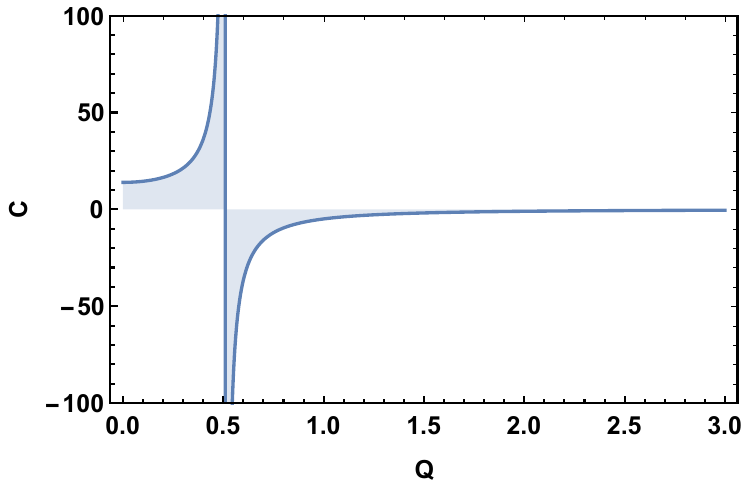}
\includegraphics[height=5cm,width=7cm]{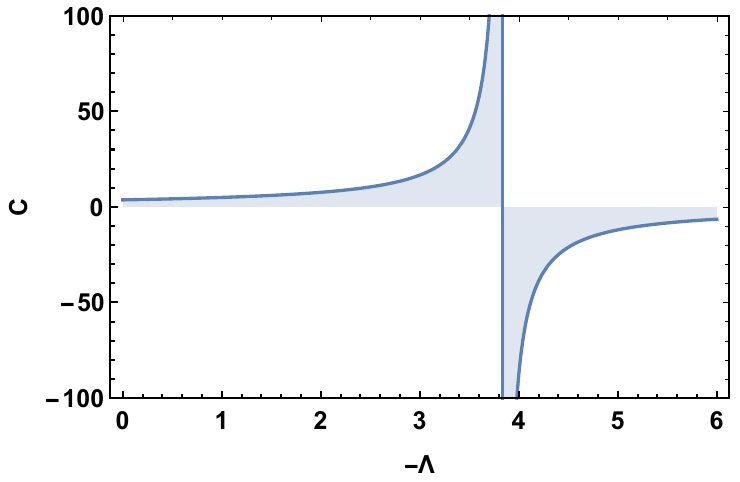}\\
(b) \hspace{8cm} (c)
\caption{Heat capacity. (a) $\Lambda=-1$ and $Q=M=r_0=1$. (b) $\Lambda=-1$, $b=0.1$ and $M=r_0=1$. (c) $b=0.1$ and $Q=M=r_0=1$.}
\label{fig14}
\end{figure}

Figure (\ref{fig13}) illustrate the behavior of the  BH entropy  $S$ as a function of the Bopp–Podolsky parameter $b$, the electric charge $Q$, and the cosmological constant $\Lambda$. In panel (a), for fixed values $\Lambda = -1$, $Q = M = r_0 = 1$, the entropy increases monotonically with increasing $b$. This trend reflects how the higher-derivative corrections induced by the Bopp–Podolsky theory lead to an effective increase in the surface area, or equivalently, the horizon structure, of the BH models. Also, the entropy in gravitational systems is typically associated with horizon area, the increasing distortion in the near-horizon geometry due to $b$ directly translates into an enhancement of entropy. In panel (b), with $b = 0.1$ and $\Lambda = -1$, the entropy rises with growing electric charge $Q$. This is expected, as a stronger electromagnetic field contributes more to the energy content and geometry of the black hole, enlarging the event horizon radius and thereby increasing entropy. Panel (c) examines how entropy varies with $-\Lambda$ (i.e., making the cosmological constant more negative), for $b = 0.1$ and $Q = M = r_0 = 1$. Here, entropy grows mildly with increasing $-\Lambda$, indicating that the enhanced AdS curvature slightly enlarges the effective horizon size, thus allowing more degrees of freedom to be encoded in the geometry.

 Figure (\ref{fig14})  focuses on the heat capacity  $C$ of the black hole and its variation under the same parameter set. In panel (a), for fixed $\Lambda = -1$ and $Q = M = r_0 = 1$, the heat capacity exhibits a non-monotonic dependence on $b$. Initially decreasing into negative values, the capacity eventually turns positive for higher $b$, suggesting a potential phase transition or stability shift. Negative heat capacity is generally a hallmark of thermodynamic instability in gravitational systems, whereas a transition to positive values signals a stable thermal configuration. This transition reflects the competition between the geometrical deformations induced by $b$ and the curvature-induced energy dispersion. In panel (b), with $b = 0.1$, $\Lambda = -1$, and $M = r_0 = 1$, increasing the electric charge $Q$ causes the heat capacity to swing more dramatically between negative and positive regions. This suggests that higher charges introduce greater fluctuations in the energy distribution, leading to a more sensitive thermal response. Aslo, panel (c) explores how the heat capacity behaves under increasing $-\Lambda$, again with fixed $b = 0.1$, $Q = M = r_0 = 1$. Here, we see a predominantly negative behavior with deeper negative peaks as $-\Lambda$ increases, indicating that  stronger AdS curvature enhances instability , possibly due to increased gravitational compression near the core.

\subsection{Second Approach}

\begin{figure}[ht]
\centering
\includegraphics[height=5cm,width=7cm]{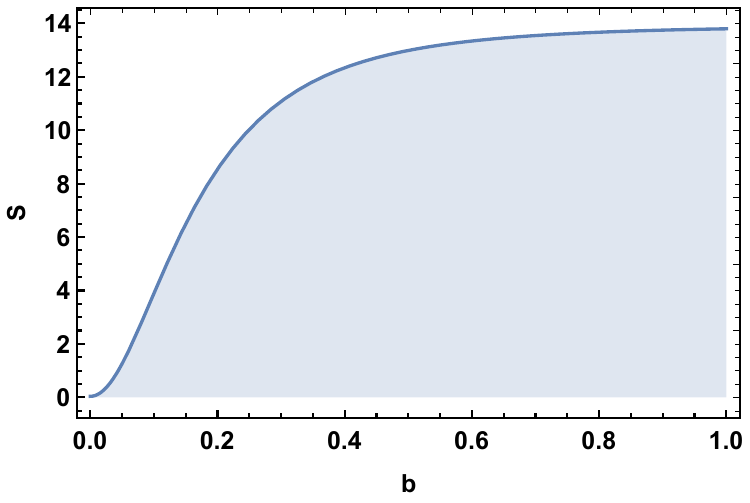}\\
(a)\\
\includegraphics[height=5cm,width=7cm]{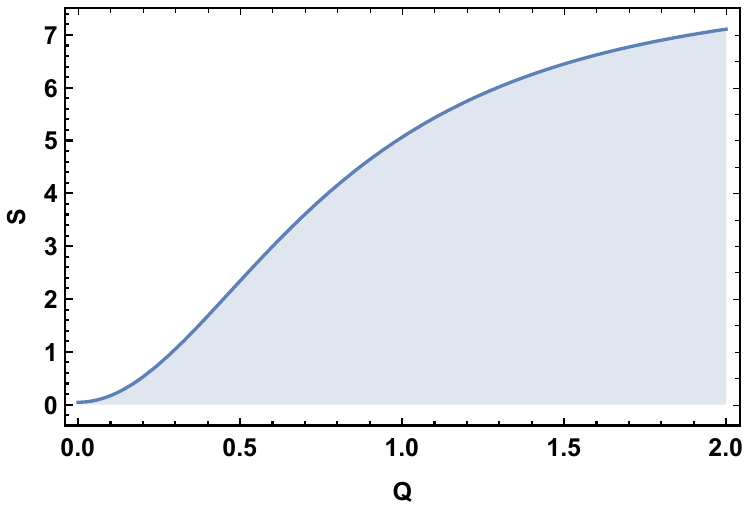}
\includegraphics[height=5cm,width=7cm]{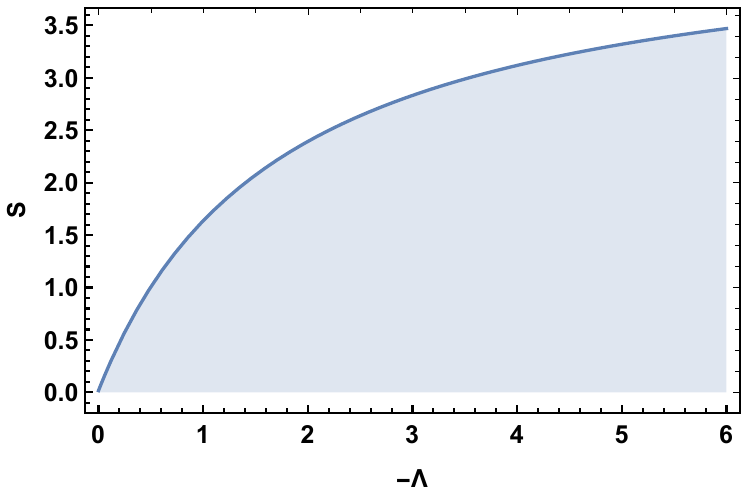}\\
(b) \hspace{8cm} (c)
\caption{Entropy. (a) $\Lambda=-1$ and $Q=M=r_0=1$. (b) $\Lambda=-1$, $b=0.1$ and $M=r_0=1$. (c) $b=0.1$ and $Q=M=r_0=1$.}
\label{fig15}
\end{figure}

Figure (\ref{fig15})  show the entropy behavior under the second-order perturbative corrections, offering a more refined and accurate representation of the thermodynamic response. In panel (a), entropy again increases with $b$, reflecting that second-order corrections amplify the role of $b$  in modifying the horizon geometry. For small values of $b$, the entropy remains nearly constant, but beyond a threshold, it rises sharply, suggesting a nonlinear sensitivity of the system to higher-order electrodynamic corrections. In panel (b), entropy is plotted against the charge $Q$ for $b = 0.1$, $\Lambda = -1$, and $M = r_0 = 1$. Panel (c), focused on the cosmological constant, again shows an increasing trend with $-\Lambda$, where the AdS curvature continues to contribute positively to the entropy. These results show that the inclusion of higher-order corrections not only refines the entropy values but also emphasizes the influence of nonlinearities inherent in the geometry.

\begin{figure}[ht]
\centering
\includegraphics[height=5cm,width=7cm]{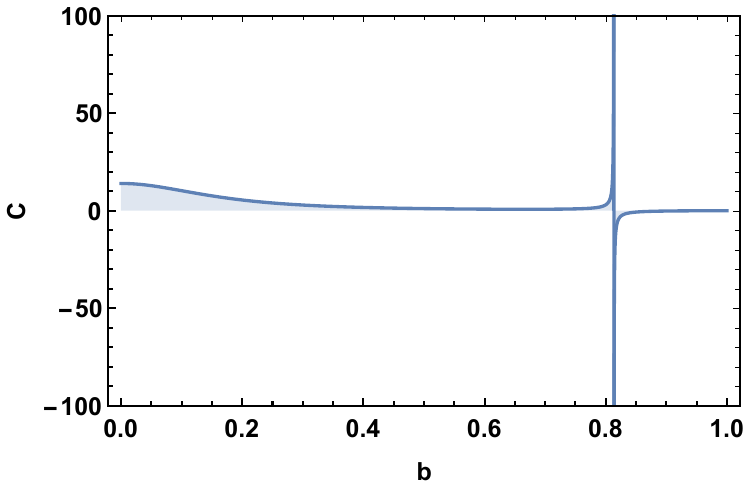}\\
(a)\\
\includegraphics[height=5cm,width=7cm]{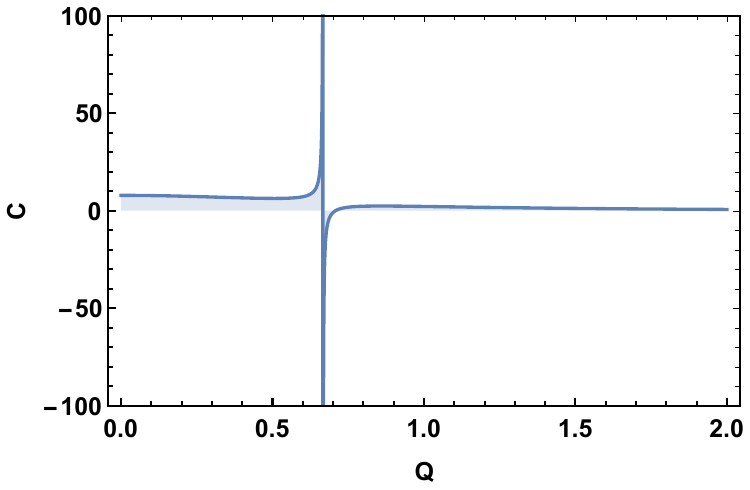}
\includegraphics[height=5cm,width=7cm]{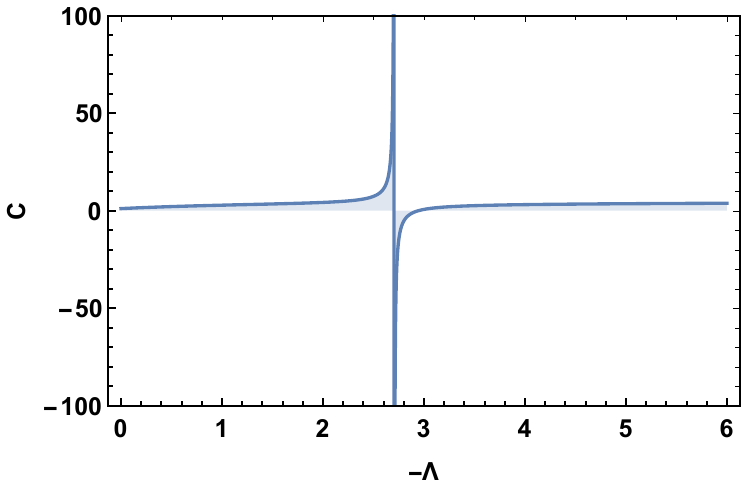}\\
(b) \hspace{8cm} (c)
\caption{Heat capacity. (a) $\Lambda=-1$ and $Q=M=r_0=1$. (b) $\Lambda=-1$, $b=0.1$ and $M=r_0=1$. (c) $b=0.1$ and $Q=M=r_0=1$.}
\label{fig16}
\end{figure}

Figure (\ref{fig16}) illustrates the second-order-corrected heat capacity, offering deeper insight into thermal stability. In panel (a), as $b$ increases (with $\Lambda = -1$ and $Q = M = r_0 = 1$), the heat capacity exhibits oscillatory behavior, with multiple sign changes suggesting possible phase transitions. Also, these transitions likely arise from the balance between higher-derivative energy contributions and the background curvature. The amplitude of fluctuations in $C$ is higher compared to Figure (\ref{fig14}), indicating that second-order terms bring out subtle instabilities missed at first order. In panel (b), for varying $Q$ and fixed $b = 0.1$, the capacity again fluctuates more wildly than in the first-order case. The system’s heat capacity appears extremely sensitive to charge perturbations, especially near certain critical charge values, which may correspond to extrema in the temperature–entropy profile. Panel (c) shows how increasing $-\Lambda$ leads to more intense and negative values of $C$, echoing the first-order trend but now with sharper transitions. In this case, the deeper wells in heat capacity signal enhanced gravitational compression from the AdS background, increasing instability. Altogether, Figure (\ref{fig16}) underscores that Bopp–Podolsky corrections and the AdS background induce rich thermal dynamics, including possible phase transitions, stability shifts, and critical behavior.

\section{Conclusion}\label{conclusion}

In this work, we have tested the thermodynamic properties of charged BTZ-type BHs within the framework of Bopp-Podolsky electrodynamics, a higher-derivative gauge-invariant extension of Maxwell’s theory. By employing a perturbative approach, we derived first- and second-order corrections to the metric and electric field, demonstrating how the Bopp-Podolsky parameter $ b $ introduces curvature-dependent modifications to the BH solution.  In a four-dimensional profile, where such corrections can lead to wormhole geometries, the (2+1)-dimensional case preserves the black hole interpretation, with deformations that decay rapidly at large distances. The resulting metric exhibits a logarithmic dependence on the radial coordinate, a feature absent in the standard charged BTZ solution, while the electric field acquires additional terms proportional to $ b^2 $, altering its asymptotic behavior. These results suggest that the lower-dimensional gravitational framework imposes stricter constraints on the possible spacetime structures arising from higher-derivative electrodynamics.  

The thermodynamic studies revealed that the Hawking temperature of the modified BH depends explicitly on the Bopp-Podolsky parameter and the cosmological constant. The temperature corrections stem from changes in the surface gravity induced by the perturbed metric, show the interplay between higher-derivative electrodynamics and BH thermodynamics. Figures (\ref{fig1}) through (\ref{fig8}) comprehensively illustrate how parameters influence under the Bopp–Podolsky parameter $b$, electric charge $Q$, cosmological constant $\Lambda$, and mass $M$, affect the geometric and thermodynamic properties of charged BTZ-type black holes under Bopp–Podolsky electrodynamics. In Figures (\ref{fig1}) and (\ref{fig5}) , the electric field $E(r)$ is analyzed. In panel (1a), $E(r)$ is plotted for different values of $b = 0.1, 0.2, 0.3$, with $\Lambda = -1$ and $Q = M = r_0 = 1$. In (1b), $Q = 1, 2, 3$ with $\Lambda = -1$, $b = 0.1$, and $M = r_0 = 1$. In (1c), $\Lambda = -1, -2, -3$ with $b = 0.1$ and $Q = M = r_0 = 1$. The same parameter combinations are used for panels (5a), (5b), and (5c) respectively. These plots show that $E(r)$ becomes stronger near the origin as $b$ and $Q$ increase, while larger magnitudes of $-\Lambda$ cause slight modifications in the field profile. Figures (\ref{fig2}) and (\ref{fig6}) present the temporal metric function $A(r)$. In panel (2a), $b = 0.1, 0.2, 0.3$ with $\Lambda = -1$, $Q = M = r_0 = 1$, (2b) uses $Q = 1, 2, 3$ with $\Lambda = -1$, $b = 0.1$, and $M = r_0 = 1$, (2c) uses $\Lambda = -1, -2, -3$ with $b = 0.1$ and $Q = M = r_0 = 1$. The same configurations are used in Figure 6 with minor extensions, where panel (6b) uses $Q = 1.0, 1.2, 1.3$ and (6c) again uses $\Lambda = -1, -2, -3$. These figures demonstrate that increasing $b$, $Q$, or $|\Lambda|$ causes deeper and more curved metric profiles near the horizon. Figures (\ref{fig3}) and (\ref{fig7}) depict the radial metric function $B(r)$. In Figure (\ref{fig3}), panel (3a) shows $b = 0.1, 0.2, 0.3$ with $\Lambda = -1$, $Q = M = r_0 = 1$, (3b) shows $Q = 1, 2, 3$ with $\Lambda = -1$, $b = 0.1$, and $M = r_0 = 1$, (3c) presents $\Lambda = -1, -2, -3$ with $b = 0.1$ and $Q = M = r_0 = 1$. In Figure (\ref{fig7}), panel (7a) uses $b = 0.05, 0.10, 0.15$, (7b) uses $Q = 1.0, 1.2, 1.3$, and (7c) uses $\Lambda = -1, -2, -3$, all with $Q = M = r_0 = 1$ and other parameters fixed as appropriate. These plots confirm that Bopp–Podolsky corrections, especially at second order, significantly influence the geometry near the black hole core. Figures (\ref{fig4}) and (\ref{fig8}) focus on the Hawking temperature $T_H$. In Figure (\ref{fig4}), panel (4a) varies $b$ with $\Lambda = -1$ and $Q = M = r_0 = 1$, (4b) varies $Q$ with $\Lambda = -1$, $b = 0.1$, and $M = r_0 = 1$, (4c) varies $\Lambda = -1, -2, -3$ with $b = 0.1$ and $Q = M = r_0 = 1$. Figure (\ref{fig8}) replicates this setup in higher resolution and more detailed plots: (8a) shows non-monotonic behavior of $T_H$ versus $b$, (8b) shows an increasing $T_H$ with $Q$, including a local minimum in the inset, and (8c) confirms that $T_H$ decreases as $-\Lambda$ increases. Overall, these figures collectively illustrate the profound impact of Bopp–Podolsky corrections and spacetime curvature on the physical properties and thermodynamic behavior of (2+1)-dimensional charged black holes.

Figures (\ref{fig13})–(\ref{fig16}) analyze the thermodynamic behavior of charged BTZ-type black holes under Bopp–Podolsky electrodynamics, focusing on the influence of the Bopp–Podolsky parameter $b$, electric charge $Q$, and cosmological constant $\Lambda$. In Figures (\ref{fig13}) and (\ref{fig15}), the entropy $S$ increases with $b$, $Q$, and $-\Lambda$, indicating that higher-derivative corrections and stronger charge or AdS curvature enlarge the horizon structure and increase the black hole's degrees of freedom. Also, Figures (\ref{fig14}) and (\ref{fig16}) show that the heat capacity $C$ exhibits non-monotonic behavior, with transitions between negative and positive values depending on the parameters, suggesting possible phase transitions and thermodynamic instabilities. In this case, these effects are more pronounced in second-order corrections (Figures (\ref{fig15}) and (\ref{fig16})), where sharper entropy growth and stronger oscillations in heat capacity reveal enhanced sensitivity to curvature and electromagnetic contributions. Aslo, the results show that Bopp–Podolsky corrections significantly affect the BH thermodynamic stability, especially near the core, with second-order effects offering deeper insight into phase behavior and geometric deformation.

In this context, Future studies could analyze numerical methods to show non-perturbative solutions and BH stability. Additionally, investigating the role of quantum fluctuations or alternative gauge-invariant extensions of electrodynamics in (2+1)-dimensional gravity. Also, our results contribute to the broader illustration of how alternative electrodynamic theories modify the BH model. Also, in lower dimensions, where exact solutions are more tractable. The persistence of BH solutions in this context, despite the presence of higher-derivative terms, underscores the unique role of dimensionality in shaping gravitational phenomena and opens new avenues for exploring quantum gravity effects in simplified settings.

\section*{Acknowledgement}
One of the authors, S.H. Dong, started this work during a research stay in China with permission from IPN, Mexico.  A.R.P. Moreira is grateful for the hospitality and support provided by the Research Center for Quantum Physics at Huzhou University.

\section*{Data Availability}

No new data were generated or analyzed in this study.

\section*{Conflict of Interests}

Authors declares there is no conflict of interests.

\section*{Funding Statement}

S.H.Dong and G.H. Sun acknowledge the partial support of the projects 20251087-SIP-IPN and 20251109-SIP-IPN, Mexico.

%%%%%%%%%%%%%%%%%%%%%%%%%%%%%%%%%%%%%%%%%%%%%%%%%%%%%%%%%%%%%%%%%%%%%%%%%%%%

\end{document}